%% file: Main.tex
\documentclass[runningheads]{llncs}
\usepackage{graphicx}
\usepackage[colorinlistoftodos, shadow, disable]{todonotes}
\usepackage{glossaries}
\usepackage{arydshln}
\usepackage[inline]{enumitem}
\usepackage[flushleft]{threeparttable}
\usepackage[utf8]{inputenc}

\usepackage[
   justification=justified,
   format=plain]{caption} 
\usepackage{subcaption}
\usepackage{csquotes}       
\usepackage[linesnumbered,ruled,vlined]{algorithm2e}
\usepackage{amsmath}
\usepackage{bm}
\usepackage{bbm}
\usepackage{sidecap}
\usepackage{changes} 
\usepackage{floatrow}
\usepackage{dsfont}
\usepackage{amsmath}
\usepackage{amsfonts} 

\usepackage{multicol}
\usepackage{nicefrac}

\newcommand{\inj}[1]{#1 }
\newcommand{\injj}[1]{#1 }

\newacronym{ML}{ML}{Machine Learning}
\newacronym{FL}{FL}{Federated Learning}
\newacronym{DLG}{DLG}{Deep Leakage from Gradients}
\newacronym{LLG}{LLG}{Label Leakage from Gradients}
\newacronym{iDLG}{iDLG}{Improved Deep Leakage from Gradients}
\newacronym{MSE}{MSE}{mean squared error}
\newacronym{IID}{IID}{independent and identically distributed}
\newacronym{SGD}{SGD}{stochastic gradient descent}
\newacronym{GAN}{GAN}{Generative Adversarial Network}
\newacronym{CNN}{CNN}{Convolutional Neural Network}
\newacronym{FCNN}{FCNN}{fully-connected neural network}
\newacronym{ASR}{ASR}{attack success rate}
\newacronym{NN}{NN}{Neural Network}

\begin{document}
\title{User-Level Label Leakage from Gradients in Federated Learning
}
%
%

\author{
Aidmar~Wainakh \and
Fabrizio~Ventola \and
Till~Müßig$^*$ \and
Jens~Keim$^*$ \and
Carlos~Garcia~Cordero \and
Ephraim~Zimmer \and
Tim~Grube \and
Kristian~Kersting \and
Max~Mühlhäuser
}
\authorrunning{A. Wainakh et al.}
%
\institute{Technical University of Darmstadt, Germany \\
\email{\emph{firstname.lastname}@tu-darmstadt.de \\  $^*$\emph{firstname.lastname}@stud.tu-darmstadt.de}  }
\maketitle              
\begin{abstract}
Federated learning enables multiple users to build a joint model by sharing their model updates (gradients), while their raw data remains local on their devices. 
In contrast to the common belief that this provides privacy benefits, 
we here add to the very recent results on privacy risks when sharing gradients.
Specifically, we investigate Label Leakage from Gradients (LLG), a novel attack to extract the labels of the users' training data from their shared gradients.
The attack exploits the direction and magnitude of gradients to determine the presence or absence of any label. 
LLG is simple yet effective, capable of leaking potential sensitive information represented by labels, and scales well to arbitrary batch sizes and multiple classes.
We mathematically and empirically demonstrate the validity of the attack under different settings. 
Moreover, empirical results show that LLG successfully extracts labels with high accuracy at the early stages of model training.
We also discuss different defense mechanisms against such leakage.
Our findings suggest that gradient compression is a practical technique to mitigate the attack.

\keywords{Label leakage \and Federated learning \and Gradient attack \and Privacy attack.}
\end{abstract}
\input{sections/1_introduction}
\input{sections/2_background}
\input{sections/3_related_work}
\input{sections/4_gradients}
\input{sections/5_methodology}
\input{sections/6_experiments}
\input{sections/8_conclusion}

\section*{Acknowledgements}
This work has been funded by the Deutsche Forschungsgemeinschaft (DFG, German Research Foundation) 251805230/GRK2050.
\emph{Fabrizio Ventola} and \emph{Kristian Kersting} acknowledge the support by the project ``kompAKI - The Competence Center on AI and Labour'', funded by the German Federal Ministry of Education and Research.
\emph{Kristian Kersting} acknowledges also the support by ``KISTRA – Use of Artificial Intelligence for Early Detection of Crimes'', a project funded by the German Federal Ministry of the Interior, Building and Community, FKZ: 13N15343, and  ``safeFBDC - Financial Big Data Cluster'', a project funded by the German Federal Ministry for Economics Affairs and Energy as part of the GAIA-x initiative, FKZ: 01MK21002K.

\bibliographystyle{splncs04}
\bibliography{__Bibliography}

\input{sections/appendix}

\end{document}

%% file: sections/1_introduction.tex
\section{Introduction}\label{sec:intro}
In an increasingly interconnected world, the abundance of data and user information has brought \gls{ML}
techniques into daily life and many services.
Arguably, the most common \gls{ML} approaches work in a centralized fashion, 
typically requiring large amounts of user data to be collected and processed by central service providers.
This data can be of sensitive nature, raising concerns about the handling of data in accordance with user expectations and privacy regulations (e.g., the European General Data Protection Regulation, GDPR).

\gls{FL} is an emerging \gls{ML} setting that allegedly enables service providers \emph{and} users to utilize the power of \gls{ML} without exposing the user's personal information.
The general principle of \gls{FL} consists of cooperating to train an \gls{ML} model in a distributed way. Users are given a model, which they can locally train with their sensitive data. 
Afterwards, users only share the model gradients of their training endeavors 
with a central server. 
The users' gradients are aggregated to establish the joint model~\cite{mcmahan2016communication}.
This general principle is currently believed to reduce the impact on users' privacy compared to the classical centralized \gls{ML} setting, since personal information does not leave the user, and sharing learning gradients does not supposedly reveal information about the user~\cite{zhu2019deep}.

However, a considerable number of recent works have shown that gradients can be exploited to reconstruct the users' training data in \gls{FL}~\cite{aono2017privacy,wang2019beyond,geiping2020inverting,zhu2020r};
while protecting the users’ ground-truth labels from possible leakage has received only limited attention~\cite{zhu2019deep,zhao2020idlg,li2020label}, mainly focusing on gradients generated from a small number of data samples (small batches) or binary classification tasks.
Label leakage, however, is a considerable risk for \gls{FL}. 
Both, \gls{FL} as well as the more superordinate setting of distributed \gls{ML} are used in many applications where labels can contain highly sensitive information.
For example, in the medical sector, hospitals employ distributed learning to collaboratively build \gls{ML} models for disease diagnosis and prediction~\cite{jochems2017developing,flores2021federated}.
\inj{In some cases, the medical data is collected directly from the patients' personal devices~\cite{daley2016using}, e.g., mobile phones~\cite{duane2017using}, where an application of \gls{FL} could introduce many potential benefits.} Building models in this and many other settings, while maintaining the users' privacy, would be crucial.
Leaking the labels of the users' data 
might disclose their diseases, which is a severe violation of privacy.
It is essential to highlight this issue and explore to what extent gradients can leak information about labels.
For this purpose, developing privacy attacks that exploit gradients is of high importance in order to foster research and development on the mitigation of respective privacy risks. 

Triggered by this, we investigate \gls{LLG}, a novel attack to extract ground-truth labels from shared gradients trained with mini-batch \gls{SGD} for multi-class classification.
\inj{\gls{LLG} is based on a combination of mathematical proofs and heuristics derived empirically.}
The attack exploits two properties that the gradients of the last layer of a neural network have:
(\textbf{P1}) The direction of these gradients indicates whether a label is part of the training batch.
(\textbf{P2}) The gradient magnitude can hint towards the number of occurrences of a label in the batch.
Here, we formalize these properties, provide their mathematical proofs, \inj{study an extended} threat model, and conduct an extensive evaluation, as follows.
\begin{itemize}
    \item We consider four benchmark datasets, namely, MNIST, SVHN, CIFAR-100, and \inj{CelebA}.
    Results show that \gls{LLG} achieves high success rate despite the datasets having different classification targets and complexity levels. 

    \item \inj{We consider two \gls{FL} algorithms, namely, FedSGD and FedAvg~\cite{mcmahan2016communication}.
    Results show that for untrained models \gls{LLG} is more effective under FedSGD, yet poses a serious threat to expose labels under FedAvg as well.}
    
    \item We study \gls{LLG} considering different capabilities of the adversary.
    Experiments demonstrate that an adversary with an auxiliary dataset, which is similar to the training dataset, can adequately extract labels with an accuracy of $>98\%$ at the early stage of the model training \inj{under the FedSGD algorithm}.
    
    \item We show that the simple LLG attack can outperform one of the state-of-the-art optimization-based attacks, \gls{DLG}~\cite{zhu2019deep}, under several settings. 
    Furthermore, \gls{LLG} is orders of magnitude faster than \gls{DLG}. 
    
    \item We also investigate the effectiveness of the attack on various model architectures including simple \gls{CNN}, LeNet~\cite{lecun1998gradient}, and ResNet20~\cite{he2016deep}.
    Results suggest that \gls{LLG} is not highly sensitive to the complexity of the model architecture. 
    
    \item We illustrate the influence of the model convergence status on \gls{LLG}.
    Findings reveal that \gls{LLG} can perform best at the early stages of training and still demonstrates information leakage in well-trained models.
    
    \item Finally, we test \gls{LLG} against two defense mechanisms: noisy gradients and gradient compression (pruning).
    Results show that gradient compression with $\ge 80\%$ compression ratio can render the attack ineffective.
\end{itemize}

In this work, we focus on the \gls{FL} and distributed \gls{ML} settings because the surface of the attacks against gradients is much wider compared with the centralized training approach.
However, \gls{LLG} can be applied in other scenarios where the gradients of \inj{a target user} are accessible by an adversary.

We proceed as follows. We start off by reviewing the background and our problem setting in Section~\ref{sec:bg}. 
Next, in Section~\ref{sec:rw}, we present related work on information leakage from gradients.
We elaborate on our findings regarding gradients properties in Section~\ref{sec:gradients}. 
The attack is then explained in Section~\ref{sec:method}. 
Before concluding, 
we present the results of our evaluation in Section~\ref{sec:exp}.

%% file: sections/2_background.tex
\section{Background}\label{sec:bg} 
In this section, we present the fundamentals of neural networks and \gls{FL}.
Then, we describe our problem setting and the threat model.

\begin{figure}[t]
    \centering  
    \includegraphics[width=0.6\columnwidth]{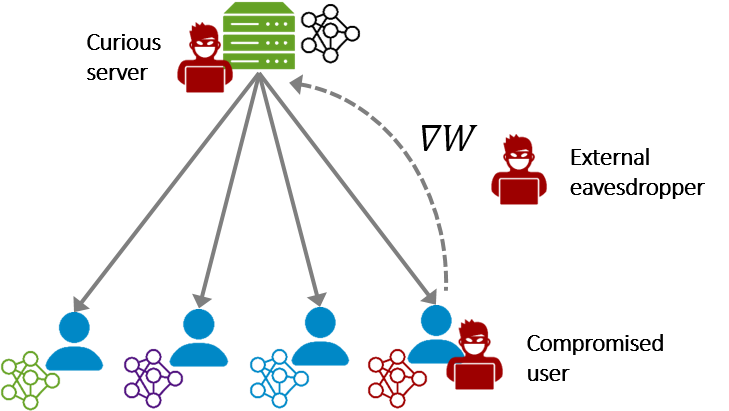}
    \caption{Federated learning overview with three potential adversary access points (in red). Gradients
    are generated by individual users and shared with a central server. 
    An adversary with access to these gradients can exploit them to estimate the presence and frequency of labels, which can be, e.g., a result of a medical imaging technique for disease prediction.}
    \label{fig:threat_overview}
\end{figure}

\subsection{Neural networks} \label{subsec:nn}
\inj{
\glspl{NN} are a subset of \gls{ML} algorithms;
an \gls{NN} is comprised of layers of nodes (neurons) including an input layer, one or more hidden layers, and an output layer.
The neurons are connected by links associated with weights~$\bm{W}$.
The \gls{NN} model can be used for a variety of tasks, e.g., regression analysis, classification, and clustering.
In the case of classification, for example, the task of the model $\hat{f}$ is to approximate the function $f(\bm{x}) = y$ where $y$ is the class label of a multidimensional data sample $\bm{x}$, e.g., an image---matrix of pixels.
To fulfill this task, the model is trained by optimizing the weights~$\bm{W}$ using a loss function~$l$ and training data consisting of input data~$\bm{x}_i: i\in [1,N]$ and corresponding labels~$y_i$ in order to solve \cite{geiping2020inverting}
\begin{equation}
    \min_{\bm{W}} \sum_{i=1}^N l_{\bm{W}} (\bm{x}_i, y_i).
\end{equation}

Minimizing the loss function can be achieved by applying one of the optimization algorithms.
Gradient descent is one of the basic optimization algorithms for finding a local minimum of a differentiable function.
This algorithm is based on gradients $\nabla \bm{W}$, which are the derivative of the loss function w.r.t. the model weights $\bm{W}$.
The core idea is to update the weights through repeated steps $t$ in the opposite direction of the gradient because this is the direction of steepest descent. 
\begin{equation}
    \bm{W}(t+1) = \bm{W}(t) - \eta \nabla \bm{W},
\end{equation}
where $\eta$ is the learning rate, which defines the step size for the model updates in the parameter space. 
An extension of gradient descent, called Minibatch Stochastic Gradient Descent is widely used for training \glspl{NN}.
This algorithm takes a batch of data samples from the training dataset to compute gradients $\nabla \bm{W}$ and, subsequently, updates the weights.
The batch size $B$ is the number of data samples given to the network for each weight update.}

\subsection{Federated learning}
\glsfirst{FL} is a machine learning setting that enables a set of $U$ users to collaboratively train a joint model under the coordination of a central server~\cite{kairouz2019advances}.
For each round $t$ of the global training process, a subset of users $K_t \ll U$ is selected to train the model locally on their data.
In particular, they optimize the model weights $\bm{W}$ based on the gradients $\nabla \bm{W}$. 
\inj{Users can take one step of gradient descent (FedSGD~\cite{mcmahan2016communication}) or multiple steps (FedAvg~\cite{mcmahan2016communication}) before sharing
the gradients $\nabla \bm{W}$ with the server.} 
The server calculates a weighted average to aggregate the gradients from the $K_t$ users, and updates the global model
\inj{
\begin{equation}
    \bm{W}(t+1) = \bm{W}(t) - \eta \sum_{k=1}^{K_t} \frac{v_k}{v} \nabla \bm{W}^k \\,
\end{equation}
}
where 
\inj{$v_k$ is the number of data samples of user $k$, and $v$ is total number of data samples.}
This process is repeated until the model potentially converges~\cite{mcmahan2016communication}.
This setting mitigates a number of privacy risks that are typically associated with conventional machine learning, where all training data should be collected, then used to train a model~\cite{kairouz2019advances}.

\subsection{Problem setting}\label{subsec:setting}
We consider a federated setting where $U$ users jointly train an \gls{NN} model \inj{for a supervised task using either the FedSGD or FedAvg algorithm~\cite{mcmahan2016communication}.
For FedSGD, the users train the model locally for one iteration on a batch of their data samples and labels.
In FedAvg, each user trains the model for several iterations (multiple batches).
We assume the users to be honest, i.e., they train the model with real data and correct labels.}
Then, the users share the gradients resulted from the local training with the server.
We assume that the model consists of $L$ layers and is trained with cross-entropy loss~\cite{Goodfellow-et-al-2016} over one-hot labels for a multi-class classification task.
For the studied attack, we focus on the gradients $\nabla \bm{W}_L$ w.r.t. the last-layer weights $\bm{W}_L$ (between the output layer and the layer before), 
\inj{where $\bm{W}_L \in \mathbb{R}^{n \times h}$: $n$ is the total number of classes and $h$ is the number of neurons in layer $L-1$.} 
The gradient vector $\nabla \bm{W}^i_L$ represents gradients connected to label $i$ on the output layer.
We note $g_i$ to refer to the sum of $\nabla \bm{W}^i_L$ elements: 
\inj{$g_i = \mathds{1}^T \cdot \nabla\bm{W}^i_L$}.

\subsection{Threat model}
\inj{We assume that an adversary applies the attack against the shared gradients of one target user.}
\inj{
The adversary analyzes the gradients to infer 
\injj{the number of label occurrences in the user's input data. 
In FedSGD, this concerns one batch, while in FedAvg, the data consists of multiple batches.} 
Thus, the more information is carried by the gradients on the labels, the higher is the privacy risk.
At the same time, the shared gradients need to reflect the training data of the users to optimize the joint model, i.e., to achieve the learning objective.
As a result, the learning objective and the depicted privacy risk are mutually related to the information carried by the gradients.
Even though it might seem as a paradox, our work is an attempt to focus on and mitigate the privacy risk imposed by gradient sharing without jeopardizing the learning objective of \gls{FL} and the model accuracy.}
Next, we define our threat model w.r.t. three aspects: adversary access point, mode, and observation.
\begin{description}[leftmargin=0cm]
    \item[Access point.] 
    The distributed nature of \gls{FL} increases the attack surface as shown in Figure~\ref{fig:threat_overview}.
    An adversary might be able to access the gradients by compromising the user's device as the gradients are calculated on the user side before sharing with the server.
    We assume that the user's device can be compromised partially, such that the adversary has no access to the training data or labels\inj{~\cite{Wei2020}.
    Such scenario can apply, for example, to several online \gls{ML} applications, where the training data is not stored but used for training on-the-fly.
    In these cases, compromising a device during or after the training phase would not grant the adversary full access to the training data, while still providing access to the model and possibly the gradients. 
    Other scenarios might exploit a vulnerability in the implementation of the network protocols/interface, such that an adversary accesses only the I/O data.}
    The server also can access the gradients of an individual user, in case no secure aggregation~\cite{bonawitz2017practical} or other protection techniques are used.
    In addition, if the connection between the server and the users is not secure, the gradients might be intercepted by an external eavesdropper. 
    
    \item[Mode.]
    We assume the adversary to act in a passive mode.
    The adversary may analyze the gradients to infer information about the users, but without hindering or deviating from the regular training protocol.
    This adversary mode is widely common in privacy attacks~\cite{pustozerovainformation,zhang2020gan,luo2020feature,zhu2019deep}, where the focus is on disclosing information rather than disturbing the system.
    
    \item[Observation.] 
    The adversary might be capable to observe different amounts of information to launch their attack.
    We consider three possibilities.
    
    \begin{enumerate}
        \item Shared gradients: the adversary has access only to the shared gradients. 
        This can apply for an external eavesdropper or an adversary with limited access to the user's device.
        
        \item White-box model: in addition to the gradients, the adversary is aware of the model architecture and parameters.
        In the case of a curious server or compromised user, the adversary might have this kind of information.
        \item Auxiliary knowledge: the adversary has access to all the aforementioned information and additionally to an auxiliary dataset.
        This dataset contains data samples of the same classes as the original training dataset.
        This is a common scenario in real-world cases, given that \glspl{NN} need a considerable amount of labeled data for training to perform accurately. 
        Labeled data is usually expensive and a typical adopted strategy is to train the model on the publicly available datasets and, eventually, fine-tuning the model on ad-hoc data. 
        Therefore, it is often easy to have access to a big part of the training data.
    \end{enumerate} 
\end{description} 

%% file: sections/3_related_work.tex
\section{Related work}\label{sec:rw}
Although the training data is not disclosed to other parties in \gls{FL}, several works in the literature showed that the data and ground-truth labels can be reconstructed by exploiting the shared gradients.
Next, we present existing 
\begin{enumerate*}[label*=(\arabic*)]
    \item data reconstruction attacks and
    \item label extraction attacks.
\end{enumerate*} 

\subsection{Data reconstruction}
Aono et al.~\cite{aono2017privacyrevisited,aono2017privacy} are the first to discuss reconstructing data from gradients and illustrate its feasibility 
on a simple \gls{NN} with a training batch of one sample.
The authors closely examined the mathematical definition of the gradients shared with the central server as proposed in~\cite{SS15}.
With the help of four examples, they showed how the relationship between the input data, which is unknown to the server, and the gradients can be exploited in order to leak at least some information about the unknown input. 
Wang et al.~\cite{wang2019beyond} moved on to generative attacks, leveraging a \gls{GAN} to reconstruct the input data in a \gls{CNN}.
Instead of training the discriminator of the \gls{GAN} on the server side with real user data, the authors observed, that locally training a shared model on each user like in the \gls{FL} setting is equivalent.
Thus, obtaining the user updates effectively yields updates to the discriminator for each user.
The generator of the GAN then is trained on the server side to generate samples indistinguishable from real user samples, which approaches the private training data.

In contrast, Zhu et al.~\cite{zhu2019deep} introduced an optimization-based attack;
the attacker generates dummy input data and output labels, then optimizes them using L-BFGS~\cite{liu1989limited} to generate dummy gradients that match the shared ones.
By that, the dummy data and labels converge to the real data and labels used by the participants in the training process.
Instead of using the euclidean distance as a cost function and L-BFGS, 
Geiping et al.~\cite{geiping2020inverting} proposed using cosine similarity and the Adam optimization algorithm.
They demonstrated that their attack is effective on trained and untrained models, also on deep networks and shallow ones.
Furthermore, they proved that the input to any fully-connected layer can be reconstructed regardless of the remaining network architecture.
\inj{Wei et al.~\cite{Wei2020} provided a framework for evaluating the reconstruction attacks and discussed the impact of multiple factors (e.g., activation and loss functions, optimizer, batch size) on the cost and effectiveness of these attacks.} 
Qian et al.~\cite{qian2020can} theoretically analyzed the limits of
~\cite{zhu2019deep} considering fully-connected \glspl{NN} and vanilla \glspl{CNN}.
They also proposed a new initialization mechanism to speed up the attack convergence.
Unlike previous approaches, Enthoven et al.~\cite{enthoven2021fidel} introduced an analytical attack that exploits fully-connected layers to reconstruct the input data on the server side, and they extend this exploitation to \glspl{CNN}.
Recently, Zhu et al.~\cite{zhu2020r} proposed a recursive closed-form attack.
They demonstrated that one can reconstruct data from gradients by recursively solving a sequence of systems of linear equations.
Overall, all the aforementioned attacks, except for~\cite{zhu2019deep}, only focus on reconstructing the input training data while overlooking the leakage of data labels, which can be of high sensitivity. 
In our research, inspired by the mathematical foundations used in these attacks, we shed more light on the potential vulnerability of label leakage in \gls{FL} and distributed learning.

\subsection{Label extraction}
While the data reconstruction attacks attracted considerable attention in the research community, a very limited number of approaches were proposed to extract the ground-truth labels from gradients. 

In the work of Zhu et al.~\cite{zhu2019deep}, the ground-truth labels are extracted as part of their optimization approach.
However, the approach requires a learning phase where the model is sensitive to the weight initialization and needs attentive hyperparamter selection.
Yet, it can be hard to converge in some cases.
Moreover, it was found to extract wrong labels frequently~\cite{zhao2020idlg} and it is effective only for gradients aggregated from a batch size $ \leq 8$~\cite{mo2020layer}. 
Actually, Zhao et al.~\cite{zhao2020idlg} proposed a more reliable analytical approach to extract the ground-truth labels by exploiting the direction of the gradients.
The authors demonstrated that the gradients of classification (cross-entropy loss) w.r.t. the last layer weights have negative values for the correct labels.
Thus, detecting the negative gradients is sufficient to extract correct labels.
However, their approach is limited to one-sample batch, which is uncommon in real-world applications of \gls{FL}, where users typically have multiple data samples and train the model on these samples (a bunch of them at least) before sharing the gradients with the server.
\inj{Wainakh et al.~\cite{wainakh2021label}, in a short paper (4 pages), introduced a basic idea to extend the attack of~\cite{zhao2020idlg} for bigger batches, however, their work lacks formalization and thorough evaluation to substantially support the validity of the approach.}
Li et al.~\cite{li2020label} proposed also an analytical approach based on the observation that the gradient norms of a particular class are generally larger than the others.
However, their approach is tailored for vertical split learning rather than \gls{FL}, and it is valid only for a binary classification task.

Overall, the existing approaches are not well generalized to arbitrary batch sizes nor number of classes.
Moreover, the influence of different model architectures on these approaches is yet to be investigated.
In our work, we take into account these issues by evaluating the \gls{LLG} attack on a variety of batch sizes and datasets with various numbers of classes, and we involve several model architectures.

%% file: sections/4_gradients.tex
\section{Gradient analysis}\label{sec:gradients}
In gradient descent optimization, the values of the gradient determine how the parameters of a model need to be adjusted to minimize the loss function.
Through an empirical analysis, we carefully 
derive two properties for the sign and magnitude of the gradients that indicate the ground-truth labels.
In this section, we formalize 
these properties, next, in Section~\ref{sec:method}, we use them as a base \inj{to launch the} attack. 

\begin{figure}[t]
    \centering  
    \includegraphics[width=0.4\columnwidth]{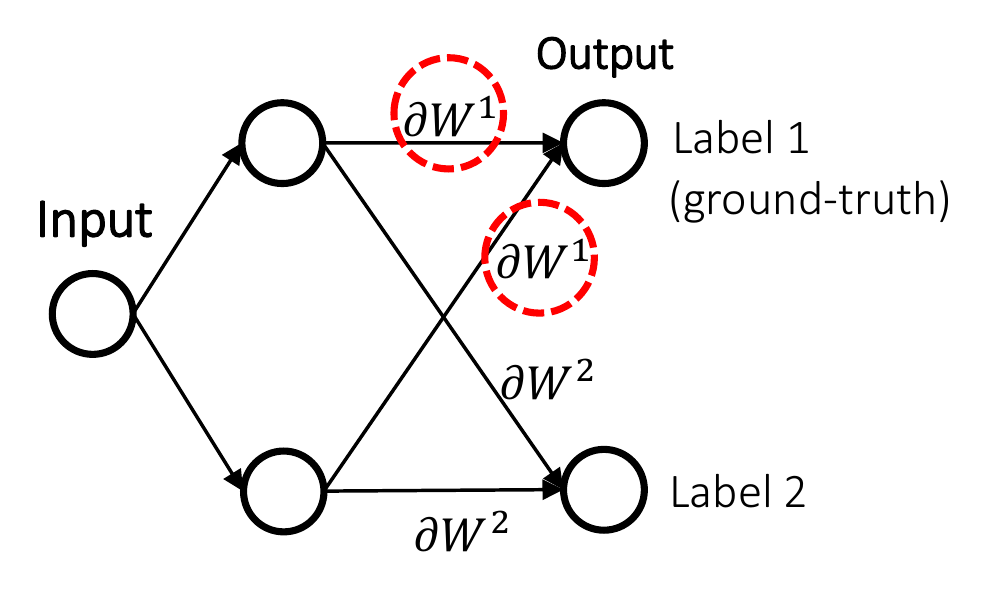}
    \caption{Graphical representation of a basic NN model and the gradients $\nabla \bm{W}^i_L$ of the last layer $L$. For simplicity, the input layer is represented by a single neuron.}
    \label{fig:model_gradients}
\end{figure} 

\begin{description}[leftmargin=0cm]
\item[Property 1.] \label{prp:1}
	\textit{For label $i$ and last layer $L$ in an \gls{NN} model with a non-negative activation function, when $ \nabla \bm{W}^i_L < 0$, label $i$ is present in the training batch on which gradient descent was applied}\footnote{\inj{This property is a generalization of the main observation in \cite{zhao2020idlg} to batches with arbitrary sizes.}}.
\end{description} 

\begin{proof}
We consider an \gls{NN} model for a classification task.
The model is trained using the cross-entropy loss over labels encoded with a one-hot encoding.
This loss function $l$ is defined as
\begin{gather}
    l(\bm{x},c) = - \ln \frac{e^{y_c}}{ \sum \nolimits_{j} e^{y_j}} \,,
    \label{eq:loss_1}
\end{gather}
where $\bm{x}$ is a multidimensional input instance and $c$ represents the ground-truth label of $\bm{x}$.
While, ${\bm{y} = [y_1, y_2, ..., y_n]}$ is the output vector of the model where each $y_i \in \bm{y}$ is the score predicted for the $i^{th}$ class, $y_c$ is the score assigned to the ground-truth label, and $n$ is the total number of classes.
A graphical representation of a simple NN model and its gradients of the last layer is depicted in Figure~\ref{fig:model_gradients}.

Given a batch size $B$, we have a set $\bm{X}$ of $B$ samples and the set of their labels $C$.
Thus, we can define a training batch as a set composed of the pairs $\{ (\bm{x}_1, c_1), \ldots, (\bm{x}_{B}, c_{B})\}$.
Therefore, we can redefine the loss function as 
the loss $l(\bm{x},c)$ of Eq.~\eqref{eq:loss_1} averaged over a batch of $B$ labeled samples
\begin{gather}
 l(\bm{X},C) = -\frac{1}{B} \sum_{k=1}^B \ln \frac{e^{y_{c(k)}}}{ \sum \nolimits_{j} e^{y_{j(k)}}} \,,
\end{gather}
where $c(k)$ is the ground-truth label for the $k^{th}$ sample in the batch, and 
$y_{c(k)}$ 
is the corresponding output score when $\bm{x}_k$ is given as input to the model.
We note that the gradient $d_i$ of the loss w.r.t. an output $y_i$ is
\begin{align}
    d_i &= \frac{\partial l(\bm{X},C)}{\partial y_{i}} = -\frac{1}{B} \sum_{k=1}^B \left( \frac{\partial \ln e^{y_{c(k)}}}{\partial y_{i}} - \frac{\partial \ln  \sum \nolimits_{j} e^{y_{j(k)}}}{\partial y_{i}} \right)
    \\
     &= - \frac{1}{B} \sum_{k=1}^B \left(\mathds{1}{(i = c(k))} - \frac{e^{y_{i(k)}}}{\sum \nolimits_{j} e^{y_{j(k)}}} \right) \,,
\end{align}
\injj{where $\mathds{1}{(\alpha=\beta)} = 1$ if $\alpha = \beta$,  $\mathds{1}{(\alpha=\beta)} = 0$ otherwise.}
\begin{align}
    d_i &= - \frac{1}{B} \sum_{k=1}^B \mathds{1}{(i = c(k))} + \frac{1}{B} \sum_{k=1}^B \frac{e^{y_{i(k)}}}{\sum \nolimits_{j} e^{y_{j(k)}}}
    \\
     &= - \frac{\lambda_i}{B} + \frac{1}{B} \sum_{k=1}^B \frac{e^{y_{i(k)}}}{\sum \nolimits_{j} e^{y_{j(k)}}} \,, \label{eq:di}
\end{align}
where $\lambda_i$ is the number of occurrences (frequency) of samples with label $i$ in the training batch.
When ${i \notin C}$, ${\lambda_i = 0}$, and ${\nicefrac{e^{y_i}}{\sum \nolimits_{j} e^{y_j}} \in (0, 1)}$, thus, ${d_i \in (0, 1)}$.
Instead, when ${i \in C}$, we have ${-\frac{\lambda_i}{B} \leq  d_i \leq 1-\frac{\lambda_i}{B}}$.
Hence, if the gradient $d_i$ is negative, we can conclude that label ${i \in C}$.
Of course, the $d_i$ value moves in this range accordingly to the status of the network weights optimization,
e.g. if ${i \in C}$ and the network performs poorly, then, $d_i$ will be closer to $-\frac{\lambda_i}{B}$.
However, the gradients $\bm{d}$ w.r.t. the outputs $\bm{y}$ are usually not calculated or shared in \gls{FL}, but only $\nabla \bm{W}$, the gradients w.r.t. the model weights $\bm{W}$. 
We write the gradient vector $\nabla \bm{W}^i_L$ w.r.t. the weights $\bm{W}^i_L$ connected to the $i^{th}$ output representing the $i^{th}$ class confidence in the output layer as follows
\begin{align}
\nabla \bm{W}^i_L &= \frac{\partial l(\bm{X},C)}{\partial \bm{W}^i_L} = \frac{\partial l(\bm{X},C)}{\partial y_i} \cdot \frac{\partial y_i}{\partial \bm{W}^i_L}\label{weights_update}
\\
&= d_i \cdot \frac{\partial ({\bm{W}^i_L}^T \bm{a}_{L-1}+b^i_L)}{\partial \bm{W}^i_L}
\\
&= d_i \cdot \bm{a}_{L-1} \label{eq:W_i} \,,
\end{align}
where $\bm{y} = \bm{a_L}$ is the activation function 
of the output layer, $b^i_L$ is the bias,
and $y_i = {\bm{W}^i_L}^T \bm{a}_{L-1}+ b^i_L$.
When non-negative activation functions (e.g. Sigmoid or ReLU) are used,  $\bm{a}_{L-1}$ is \inj{non-negative.}
Consequently, $\nabla \bm{W}^i_L$ and $d_i$ have the same sign.
Considering Eq.~\eqref{eq:di}, we conclude that negative $\nabla \bm{W}^i_L$ indicates that the label $i$ is present in the ground-truth labels set $C$ of the training batch.
\injj{However, a present label can have a positive gradient according to the value of $d_i$ as discussed earlier.}
\end{proof}

\begin{description}[leftmargin=0cm]

\item[Property 2.] \label{prp:2}
\textit{In untrained models, the magnitude of the gradient \inj{${g_i = \mathds{1}^T \cdot \nabla\bm{W}^i_L}$} is \inj{approximately} proportional to the number of occurrences $\lambda_i$ of label $i$ in the training batch.} 
\end{description}

\begin{proof} 
Based on Eq.~\eqref{eq:W_i}, we have
\begin{align}
g_i = \mathds{1}^T \cdot \nabla\bm{W}^i_L &= d_i \left(  \mathds{1}^T \cdot \bm{a}_{L-1} \right) \,.
\end{align}
We substitute $d_i$  with its expression from Eq.~\eqref{eq:di} as follows
\begin{align}
g_i &= \left(- \frac{\lambda_i}{B} + \frac{1}{B} \sum_{k=1}^B \frac{e^{y_{i(k)}}}{\sum \nolimits_{j} e^{y_{j(k)}}}\right) \left( \mathds{1}^T \cdot \bm{a}_{L-1} \right) \,. \label{eq:g_i} 
\end{align}
\inj{
When $\sum_{k=1}^B \frac{e^{y_{i(k)}}}{\sum \nolimits_{j} e^{y_{j(k)}}}$ is close to zero, we can write
\begin{align} \label{eq:gi_m}
g_i \approx - \frac{\lambda_i}{B} \left( \mathds{1}^T \cdot \bm{a}_{L-1} \right) \,, 
\end{align}
thus, $g_i$ is proportional to $\lambda_i$. 
We denote $m$ to be
\begin{align} \label{eq:impact_def}
m = - \frac{\mathds{1}^T \cdot \bm{a}_{L-1}}{B} \,, 
\end{align}
therefore, $g_i \approx \lambda_i m$.
We define the parameter \emph{impact} $m$ as \emph{the change of the gradient value caused by a single occurrence of a label in the training batch}.
This value is negative and constant across labels, thus, label-agnostic. 

However, for an untrained model, the value of $\sum_{k=1}^B \frac{e^{y_{i(k)}}}{\sum \nolimits_{j} e^{y_{j(k)}}}$ strongly depends on the model weight initialization. 
The prediction score $y_i$ can be randomly distributed around an uniform random guess $P=\nicefrac{1}{n}$, which the more classes exist in the dataset, the lower is its value, thus, the aforementioned summation goes closer to zero.
In some cases, $y_i$ might be notably high, although the label $i$ is not present in the training batch.
This comes as a result of misclassification and leads to a positive shift in the gradient values.
We call this shift \emph{offset} $s$, and based on Eq.~\eqref{eq:g_i}, we can write
\begin{align}\label{eq:offset_def}
s_i &= \left( \frac{1}{B} \sum_{k=1}^B \frac{e^{y_{i(k)}}}{\sum \nolimits_{j} e^{y_{j(k)}}}\right) \left( \mathds{1}^T \cdot \bm{a}_{L-1}\right) \,.
\end{align}
This offset value varies from a label to another, so it is a label-specific value.
Using our defined parameters impact $m$ and offset $s_i$, we can reformulate Eq.~\eqref{eq:g_i} as follows $g_i = \lambda_i m + s_i$.
From this equation, 
it follows easily that the number of occurrences~$\lambda_i$ of label~$i$ can be derived from the parameters $m$, $s_i$, and $g_i$.
}

\begin{figure*}[t]
    \centering    
    \begin{subfigure}[b]{0.336\textwidth}
        \includegraphics[trim={0.5cm 0 0.2cm 0}, clip, width=\textwidth]{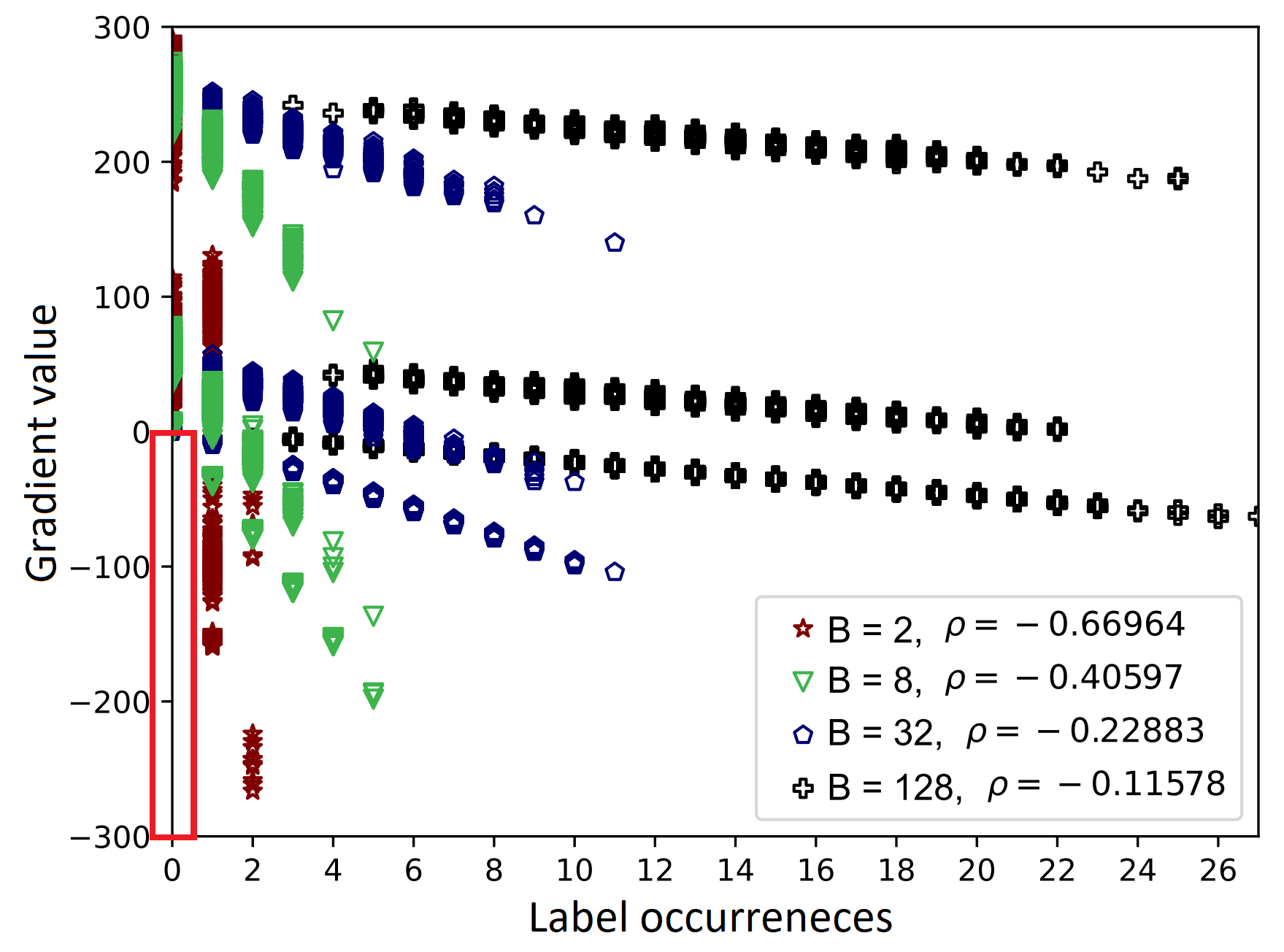}
        \caption{Gradients distribution}
        \label{fig:mnist_grad_dist}
    \end{subfigure}
    \begin{subfigure}[b]{0.325\textwidth}
        \includegraphics[trim={0.9cm 0 1.2cm 0}, clip, width=\textwidth]{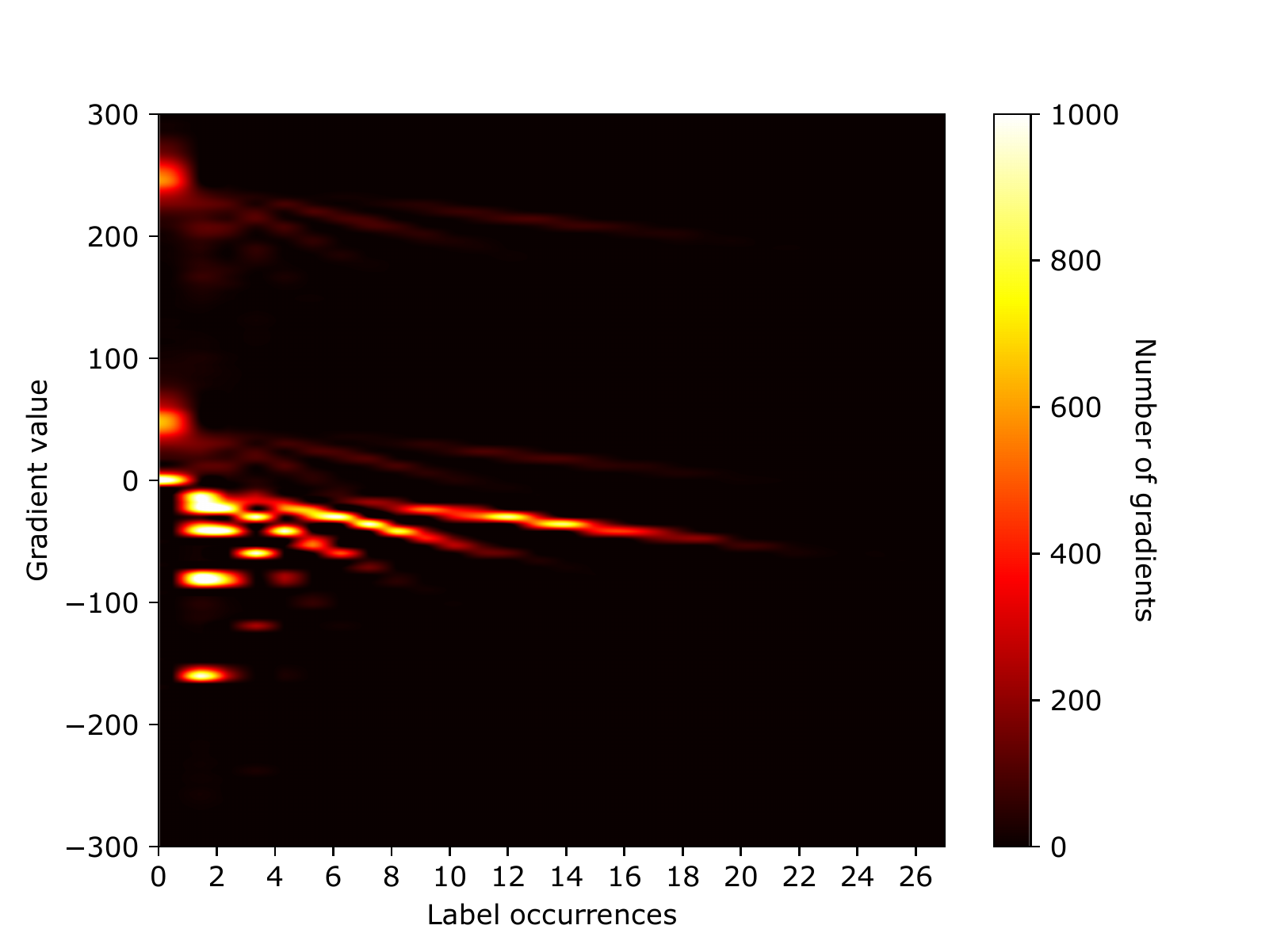}
        \caption{Gradients heatmap}
        \label{fig:mnist_heatmap}
    \end{subfigure}
    \begin{subfigure}[b]{0.317\textwidth}
        \includegraphics[trim={3cm 0 0.2cm 0}, clip, width=\textwidth]{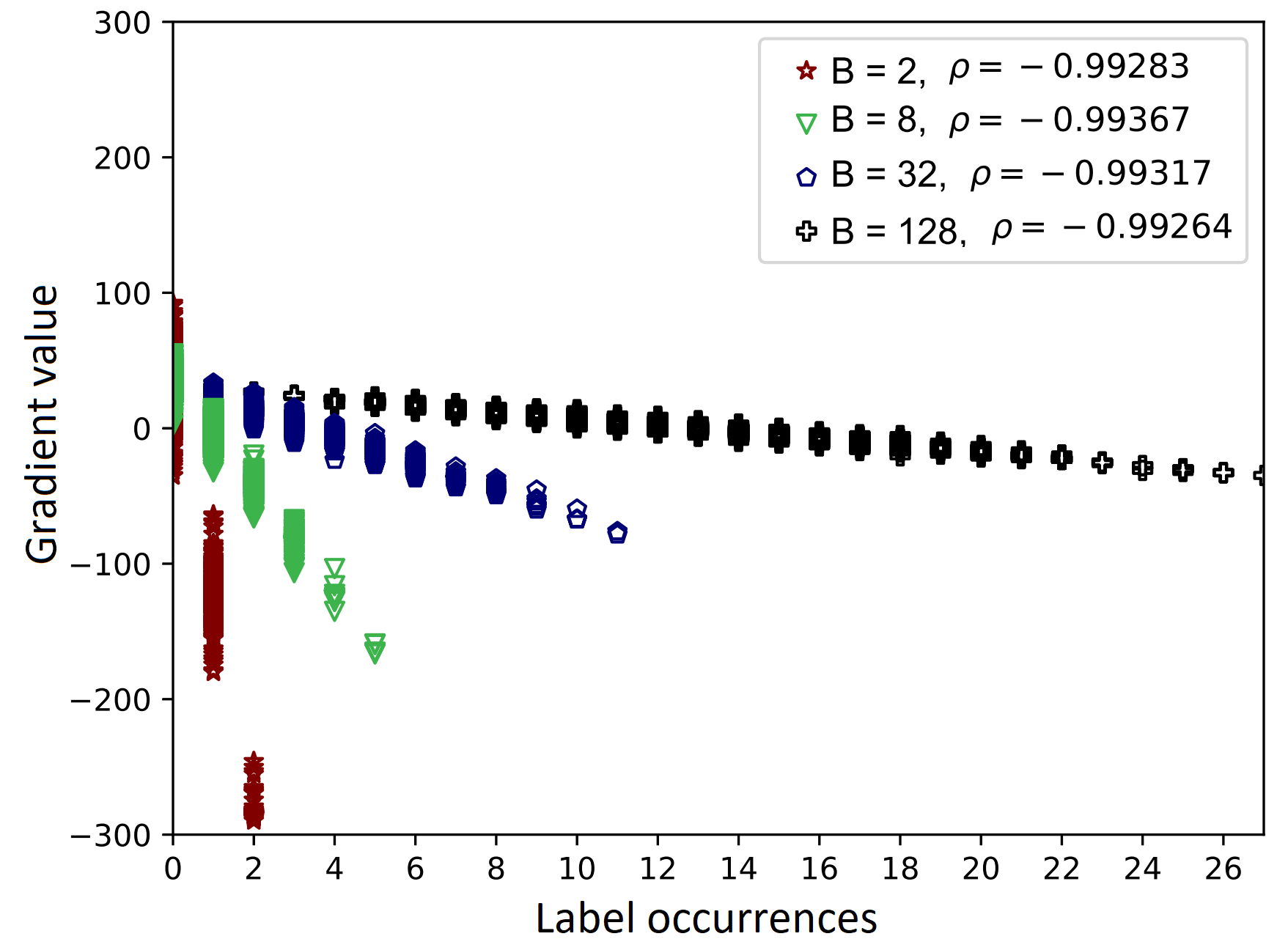}
        \caption{After calibration}
        \label{fig:mnist_after}
    \end{subfigure}
    
    \caption{\footnotesize Distribution of gradients obtained from a randomly initialized CNN on a batch of samples of MNIST varying the batch size in $\{2, 8, 32, 128\}$: (a) the distribution shows the correlation between the gradients and the label occurrences, (b) heatmap shows that the majority of the gradients have negative values when the corresponding label is present in the batch, (c) gradients after calibration exhibit a more prominent correlation with label occurrences. 
    Given this strong correlation, it is possible to accurately estimate the label occurrences in the training batch basing on the gradient values.} 
    \label{fig:grad_dist}
\end{figure*}

\end{proof}

To demonstrate the two gradient properties, we randomly initialize the weights of a CNN composed by three convolutional layers. 
Then, we check the gradients $g_i$ by evaluating the network on a batch of samples
taken from the MNIST dataset~\cite{lecun1998gradient}, \injj{which contains $10$ classes.}
We repeat the experiment~$1,000$ times with different batch sizes $B \in \{2,8,32,128\}$.
Figure~\ref{fig:grad_dist}~(a) depicts the distribution of the resulting gradients, \injj{where each data point represents the gradient value of one label in one experiment.}
The y-axis shows the gradient values and the x-axis represents the number of occurrences for the corresponding label $i: \forall i \in [1,n]$.

We can see that \inj{there are no negative gradients at $\lambda_i = 0$ (framed in red), i.e.,} the negative gradients always correspond to an existing label in the batch $\lambda_i > 0$, which confirms Property~1. 
For all the batch sizes, we notice that the values of the gradients decrease consistently with the increase of the occurrences.
This, in turn, confirms Property~2 
and our definition of the impact parameter.
We also observe that the decrease of gradient values is roughly constant regardless of the label.
This confirms the impact being label-agnostic, as we described earlier.
Furthermore, we notice that the magnitude of the impact is negatively correlated with the batch size.
Meaning, the more samples are present in a batch, the smaller are the changes of the gradients for a different number of occurrences. 
This is also clear from the definition of impact in Eq.~\eqref{eq:impact_def}.
\inj{We also can see that there are positive gradients that correspond to $\lambda_i > 0$.
The positive value of these gradients is mainly caused by the offset $s_i$ defined in Eq.~\eqref{eq:offset_def}.
To illustrate their ratio, we depict a heatmap in Figure~\ref{fig:grad_dist}~(b).}
We observe that only part of the gradients (18\%) are positive, i.e., shifted by the offset, while the majority of the gradients have negative values when the corresponding labels are present in the batch. 
\inj{In Section~\ref{sub:parameters}, we describe our methods to estimate the offset and elaborate on Figure~\ref{fig:grad_dist}~(c).}

%% file: sections/5_methodology.tex
\section{Label extraction}\label{sec:method}
In this section, we present \glsfirst{LLG}, to extract the ground-truth labels from shared gradients.
We first introduce different methods to estimate the attack parameters, impact and offset.
Then, we explain the attack.

\subsection{Attack parameters estimation}\label{sub:parameters}
In the light of different threat models, we \inj{empirically} developed several \inj{heuristic} methods to estimate the impact and offset.
\begin{description}[leftmargin=0cm]
    \item[Shared gradients.] 
        In this scenario, the adversary has access only to the shared gradients. 
        As mentioned earlier, the impact refers to the change in the value of the gradients corresponding to one occurrence of a label.
        \inj{Our intuition is that a good estimation for the impact is obtained by averaging the gradients 
        \injj{over the number of data samples $|\bm{D}|$ used by a user in a training round.
        For FedSGD, $|\bm{D}| = B$ the batch size, while for FedAvg, $|\bm{D}| = \gamma.B$, where $\gamma$ is the number of local iterations (batches).}
        Based on Property~1, 
        we know that all negative gradients are indeed indicating existing labels in the training \injj{samples.}
        Therefore, we average only the gradients with negative values.
        Consequently, this average is an underestimation since some gradients may be positive because they are shifted with an offset.
        We empirically observed that multiplying by a factor that depends on the total number of classes $n$ is a good additive correction, precisely, we multiply by $(1+1/n)$.}
        Thus, we estimate the impact $m$ as follows
        \begin{equation}\label{eq:impact1}
        m = \frac{1}{\injj{|\bm{D}|}}\sum_{i:g_i<0}^n(g_i)\left(1+\frac{1}{n}\right) \,. 
        \end{equation}
        For this threat model, the offset $s_i$ cannot be estimated, thus, considered to be zero in the attack. 
        
    \item[White-box model.] 
        When the adversary additionally has access to the model architecture \injj{and parameters,} they can use it to generate more gradients and gain more insights about the behavior of the gradients in this model.
        Consequently, better estimations for the impact and offset can be achieved.
        \inj{The approximation in Eq.~\eqref{eq:gi_m} indicates that the impact $m$ can be estimated if the gradient $g_i$ and number of occurrences $\lambda_i$ are known, regardless the quality of the input data.
        Thus, dummy data samples, e.g., dummy images of zeros (black), ones (white), or random pixels, can be used to generate $g_i$ under known $\lambda_i$.}
        More precisely, we form a collection of dummy batches, each batch contains data samples assigned to one label $i$.
        For impact estimation, we pass these batches to \injj{a shadow} model \injj{(a copy of the original model),} one at a time, and calculate the average $\bar{g_i}$ for all the batches corresponding to each label $i \in [1,n]$.
        Then, we average over all classes $n$ and the batch size $B$ as follows
        \begin{equation}\label{eq:impact2}
        m = \frac{1}{nB}\sum_{i=1}^n(\bar{g_i})\left(1+\frac{1}{n}\right)\,. 
        \end{equation} 
        As mentioned earlier, we assume the offset $s_i$ to be an approximation of misclassification penalties, when the model mistakenly predicts $i$ to be ground-truth.
        \inj{These penalties are mainly related to the status of the model weights, which can be biased to specific classes.}
        Based on this intuition, we estimate the offset $s_i$ by passing batches full of other labels ${\forall j \in [1,n]\!: j \neq i}$, each batch full of one label, one batch per run.
        We repeat this for various batch sizes, in total of $z$ runs.
        In these runs, the gradients of label $i$ reflect to some extent the misclassification penalties.
        Therefore, we calculate the mean of these gradients to be our estimated offset, thus, we have
        \begin{equation}\label{eq:offset}
        s_i = \frac{1}{z}\sum_{k=1}^z(g_{i_k}) \,.       
        \end{equation}

    \item[Auxiliary knowledge.] 
        In this scenario, the adversary has access to the shared gradients, model, and auxiliary data that contains the same classes as the training dataset. 
        Here, the adversary can follow the same methods of the white-box scenario, however, using real input data instead of dummy data.
        This in turn is expected to yield better estimations for the impact and offset.
        \inj{The goodness of the auxiliary data, i.e., the similarity of the content and class distribution to the original dataset, might play a role in the quality of the estimations.
        This aspect can be investigated in further research.}

        \quad To demonstrate the quality of our offset estimation, we calibrate the gradients of Figure~\ref{fig:grad_dist}~(a) by subtracting the estimated offset and plot the results in
        Figure~\ref{fig:grad_dist}~(c). 
        We can see how the gradients become mainly negative and strongly correlated with the label occurrences.
        \inj{To measure the correlation, we use the Pearson correlation coefficient ${-1 \leq \rho \leq 1}$~\cite{benesty2009pearson}, which yields, for all the studied batch sizes, values of ${|\rho| > 0.99}$}.
        The calibration process \inj{mitigates} the effect of the offset and makes the gradient values more consistent, thus, easier to be used for extracting the labels.
\end{description} 
 
\subsection{Label leakage from gradients attack}\label{sub:algo}
\gls{LLG} extracts the ground-truth labels from gradients by exploiting Property~1 and~2. 
The attack consists of three main steps summarized in Algorithm~\ref{algo:llg}. 
\begin{enumerate}
    \item We start with extracting the labels based on the negative values of the gradients (Property~1). 
    Thus, the corresponding label of each negative gradient is added to the list of the extracted labels $E$.
    As Property~1 
    holds firm in our problem setting, we can guarantee 100\% correctness of the extracted labels in this step.
    \inj{As preparation for the next step,} every time we add a label to $E$, we subtract the impact from the corresponding gradient following Property~2 (Lines~1-5).
    \item We calibrate the gradients by 
    subtracting the offset. 
    In case the offset is not estimated, it is considered to be zero.
    This step 
    increases the correlation between the gradient values and label occurrences, which facilitates better label extraction based on these values (Line~7).
    \item After calibration, the minimum gradient value (negative with maximum magnitude) is more likely corresponding to a label occurred in the batch 
    (see Figure~\ref{fig:grad_dist}~(c)).
    Therefore, we select the minimum and add the corresponding label to the extracted labels.
    We repeat Step (3) until the size of the extracted labels list $E$ matches the number of data samples $D$ used to generate the gradients.
    Assuming that $|D|$ is known or can be guessed by the adversary (Lines~8-11).
\end{enumerate}
Finally, the output of the \gls{LLG} attack is the list of extracted labels $E$, precisely, the labels existing in the batch and how many times they occur.

\begin{algorithm}[t!]
    \SetAlgoLined
    \KwData{\injj{$\bm{G}=[g_1,..,g_n]$:~vector} of gradients,  $m$:~impact, \injj{${\bm{S}=[s_1,..,s_n]}$:~vector} of offsets, 
    $\bm{D}$:~data samples used to generate $\bm{G}$.
    } 
    \KwResult{$E$: list for extracted labels.}
    \For{$g_i \in \bm{G}$}{ 
        \If{$g_i < 0$}{
            append $i$ to $E$\;  
            $g_i \leftarrow g_i - m$\;             
        }
    }
    \injj{$\bm{G} \leftarrow \bm{G} - \bm{S}$\; }      
    \While{$|E| < |\bm{D}|$}{
        Select $g_i: g_i = min(\bm{G})$\;
        append $i$ to $E$\;  
        $g_i \leftarrow g_i - m$\;             
    }
\caption{Label Leakage from Gradients}
 \label{algo:llg}
\end{algorithm}

%% file: sections/6_experiments.tex
\section{Empirical evaluation}\label{sec:exp}
We evaluate the effectiveness of \gls{LLG} with varying settings including: different \inj{\gls{FL} algorithms,} threat models, model architectures, and model convergence statuses.
We also test the robustness of \gls{LLG} against two defense mechanisms, namely, noisy and compressed gradients. 
For the sake of simplicity, we refer to $g_i = \mathds{1}^T \cdot \nabla\bm{W}^i_L$ as the gradient of label $i$ in the rest of this section. 
Next, we describe the experimental setting, then we discuss our results.
The source code of the experiments can be found in \url{https://github.com/tklab-tud/LLG}.

\subsection{Experimental setup}
\begin{description}[leftmargin=0cm]
    \item[Default model.]
    We use a CNN model with three convolutional layers (see Appendix, Table~\ref{tbl:cnn_arch}) as our default model for a classification task. 
    The activation function is Sigmoid, and we use SGD as an optimizer with learning rate~$0.1$ and cross-entropy as loss function.
    \inj{We use batches of varying sizes $B = 2^k: k \in [0,7]$.
    When applying the attack for FedSGD, we feed the model with one batch,
    and we use $\gamma = 10$ batches for FedAvg.}
    The label distribution in a batch can be \emph{balanced} or \emph{unbalanced}.
    For balanced data, the samples of the batch are selected randomly from the dataset.
    For unbalanced data, we select 50\% of the batch samples from one random label~$i$ and 25\% from another label~$j$.
    The remaining 25\% of the batch is chosen randomly.
    We initialize the model with random weights and repeat each experiment 100 times, then report the mean values for analysis and discussion.
\item[Datasets.]
    We conduct our experiments on four widely used benchmark datasets:
        MNIST~\cite{lecun1998gradient} consists of 70,000 grey-scale images for handwritten digits, with 10 classes in total.
        SVHN~\cite{netzer2011reading} has 99,289 color images of house numbers with 10 classes. 
        CIFAR-100~\cite{krizhevsky2009learning} contains 60,000 color images with 100 classes.
        And \inj{CelebA~\cite{liu2015deep} is a facial attributes dataset with 202,599 images.
        In our experiments, we consider only the hair color attribute with 5 classes.}

\item[Threat model.]
    \inj{We assume the users to train the model on real data and correct labels.
    The adversary has access to the shared gradients of only one target user.}
    We consider three different scenarios for the observation capabilities of the adversary (see Section~\ref{subsec:setting}).
    Based on these scenarios, the estimation of the impact and offset parameters differs (see Section~\ref{sub:parameters}), while the same attack applies for all.
    We refer to the application of the attack under these different scenarios as follows:
    \begin{enumerate}
        \item \textit{\gls{LLG}} for accessing only the shared gradients scenario. 
        \item \textit{\gls{LLG}*} for the white-box model, \inj{where we employ various dummy images to estimate the impact and offset.
        Empirically, we observed the dummy images with which the attack achieves better performance on each dataset.
        This resulted in using zeros (black) images for MNIST, random pixels for SVHN, ones (white) for CIFAR, and zeros (black) for CelebA.}
        \item \textit{\gls{LLG}+} for auxiliary knowledge, \inj{where it is assumed that the adversary has access to auxiliary data that contains $10$ batches of images from each class.}
    \end{enumerate}

\item[Metrics.]
    To measure the attack effectiveness, we use 
    the \inj{\gls{ASR} metric~\cite{Wei2020},} which is expressed as the ratio of the correctly extracted labels over the total number of the extracted labels. 
    We also employed the Hellinger distance~\cite{cramer1999mathematical} to measure the distance between the distribution of the extracted labels and the ground-truth.
    However, during our experiments, we observed that both aforementioned metrics yielded very similar measurements, therefore, we present our results only with the \gls{ASR} metric.
\item[Baselines.]
    We compare \gls{LLG} with two baselines.
    First, the DLG attack~\cite{zhu2019deep}, which aims to reconstruct the training data and labels using an optimization approach. 
    For our experiments, we run DLG for $100$ iterations and focus only on the label reconstruction results.
    \inj{We used the DLG implementation provided by Zhao et al.~\cite{zhao2020idlg}\footnote{https://github.com/PatrickZH/Improved-Deep-Leakage-from-Gradients}.}
    Second, we consider a uniform distribution-based random guess as a baseline.
    An adversary without any shared gradients might partially succeed in guessing the existing labels frequency
    by assuming that the labels distribute uniformly, especially in the case of large balanced batches.
    The random guess serves as a risk assessment curve. 
    Having any attack performing better than the random guess means that there is information leakage.
\end{description}

\subsection{Attack success rate}

\begin{figure*}[t]
    \centering  
    \begin{subfigure}[b]{\textwidth}
    \centering
        \includegraphics[width=0.45\textwidth]{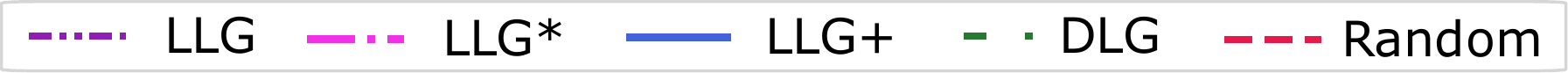}
    \end{subfigure}

        \begin{subfigure}[b]{0.26\textwidth}
        \includegraphics[trim={0 0 0.4cm 0}, clip, width=\textwidth]{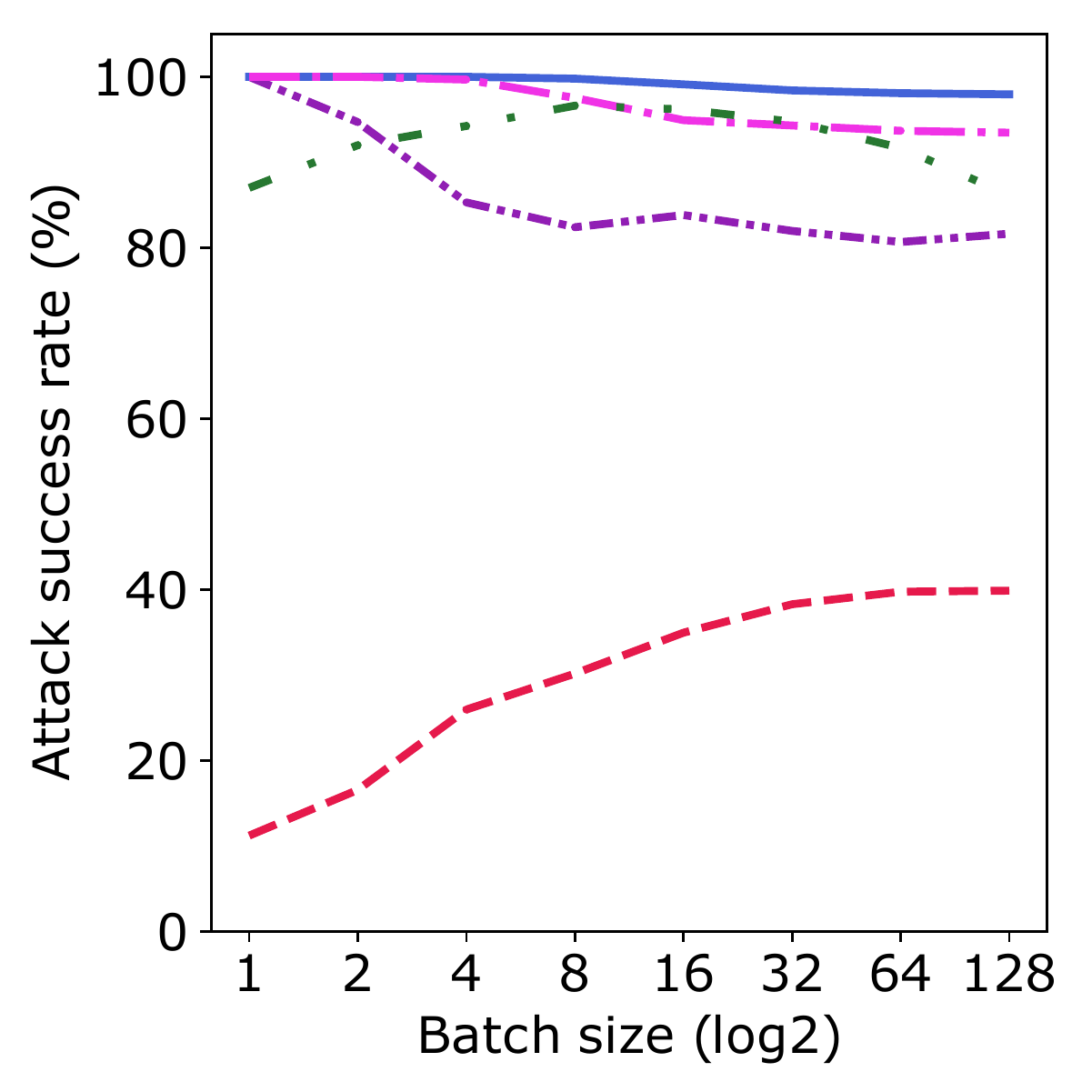} 
        \caption{MNIST - FS} 
        \label{fig:mnist_accuracy_noniid_sgd}
    \end{subfigure}   
    \begin{subfigure}[b]{0.235\textwidth}
        \includegraphics[trim={1.1cm 0 0.4cm 0}, clip, width=\textwidth]{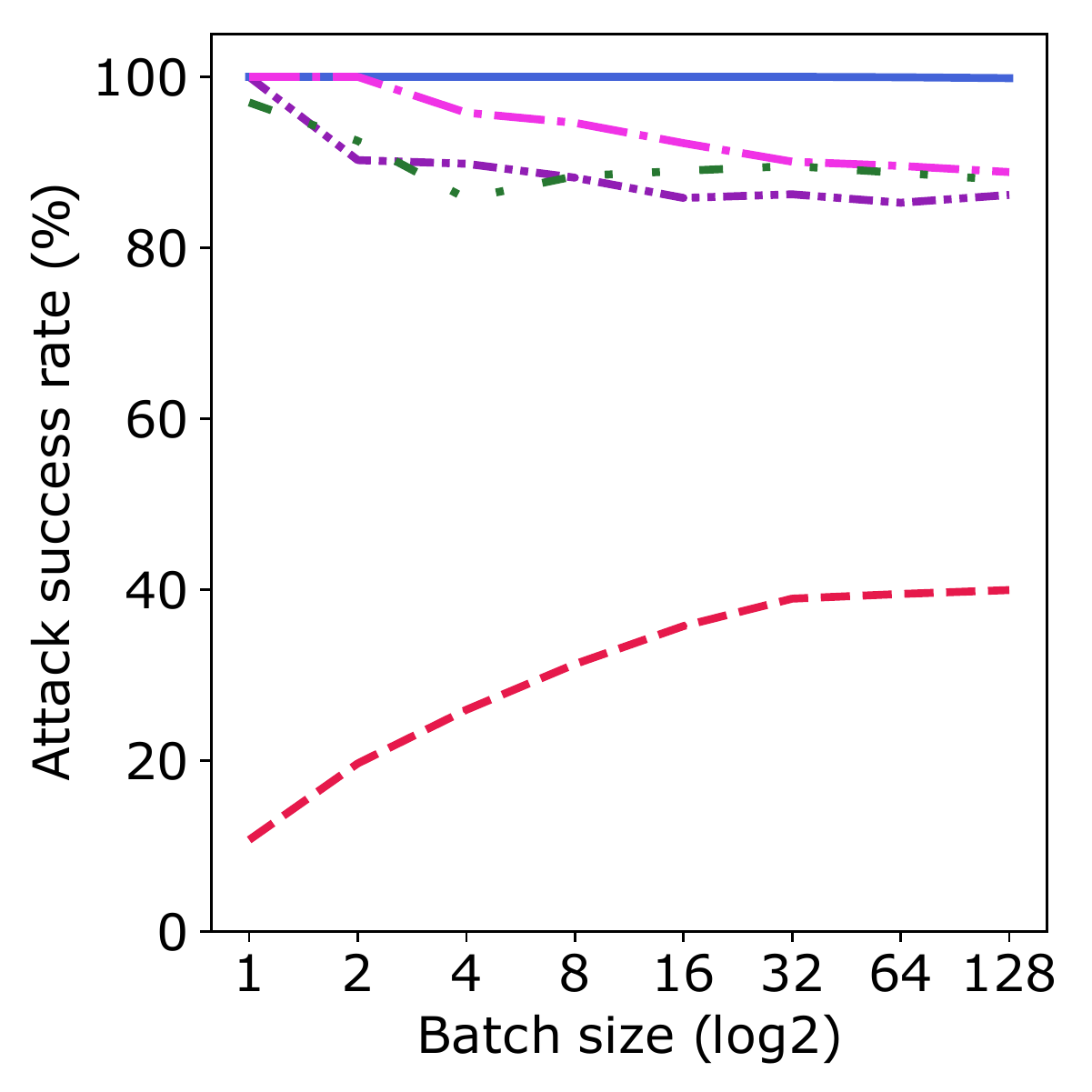} 
        \caption{SVHN - FS} 
        \label{fig:svhn_accuracy_noniid_sgd}
    \end{subfigure} 
    \begin{subfigure}[b]{0.235\textwidth}
        \includegraphics[trim={1.1cm 0 0.4cm 0}, clip, width=\textwidth]{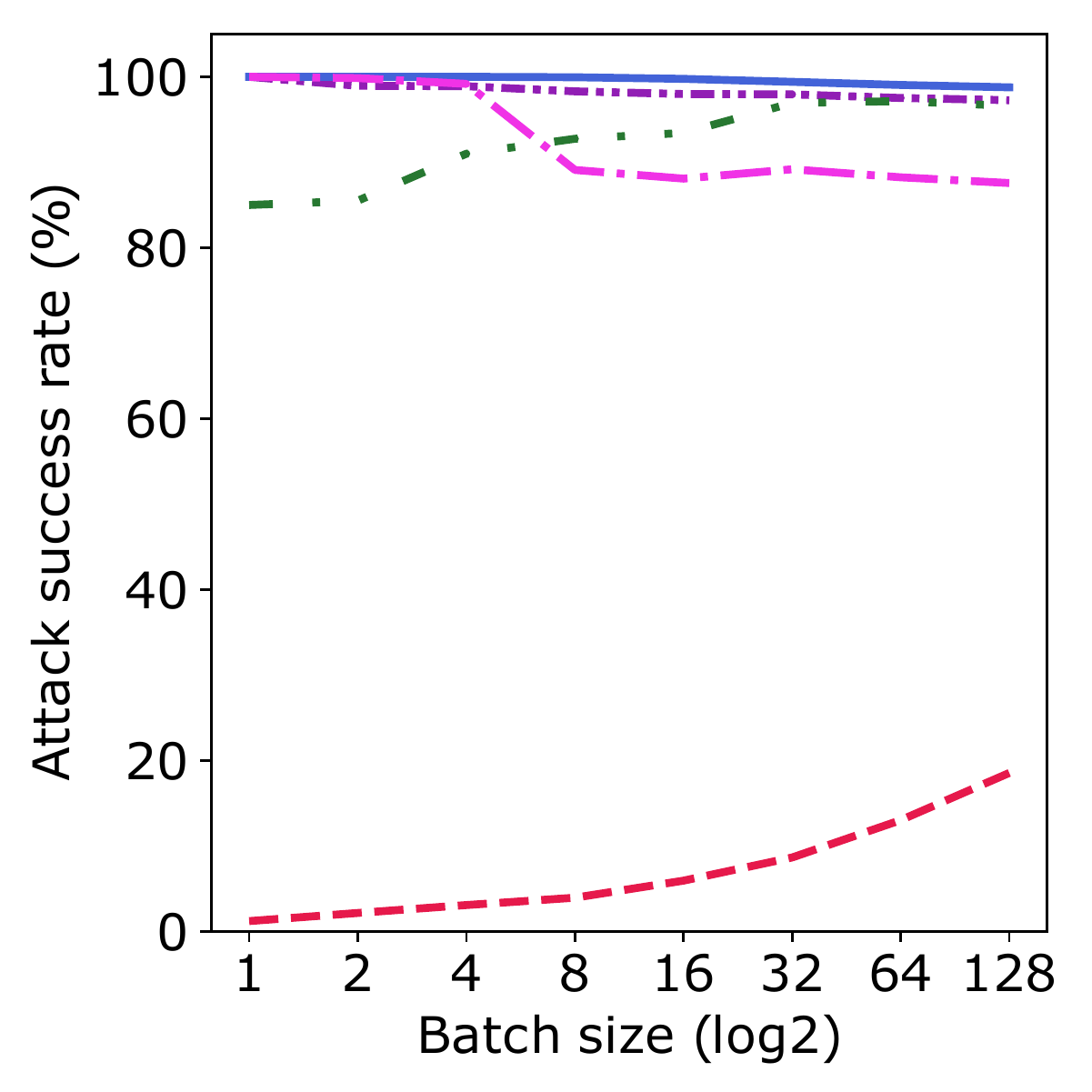} 
        \caption{CIFAR-100 - FS} 
        \label{fig:cifar_accuracy_noniid_sgd}
    \end{subfigure}  
     \begin{subfigure}[b]{0.235\textwidth}
        \includegraphics[trim={1.1cm 0 0.4cm 0}, clip, width=\textwidth]{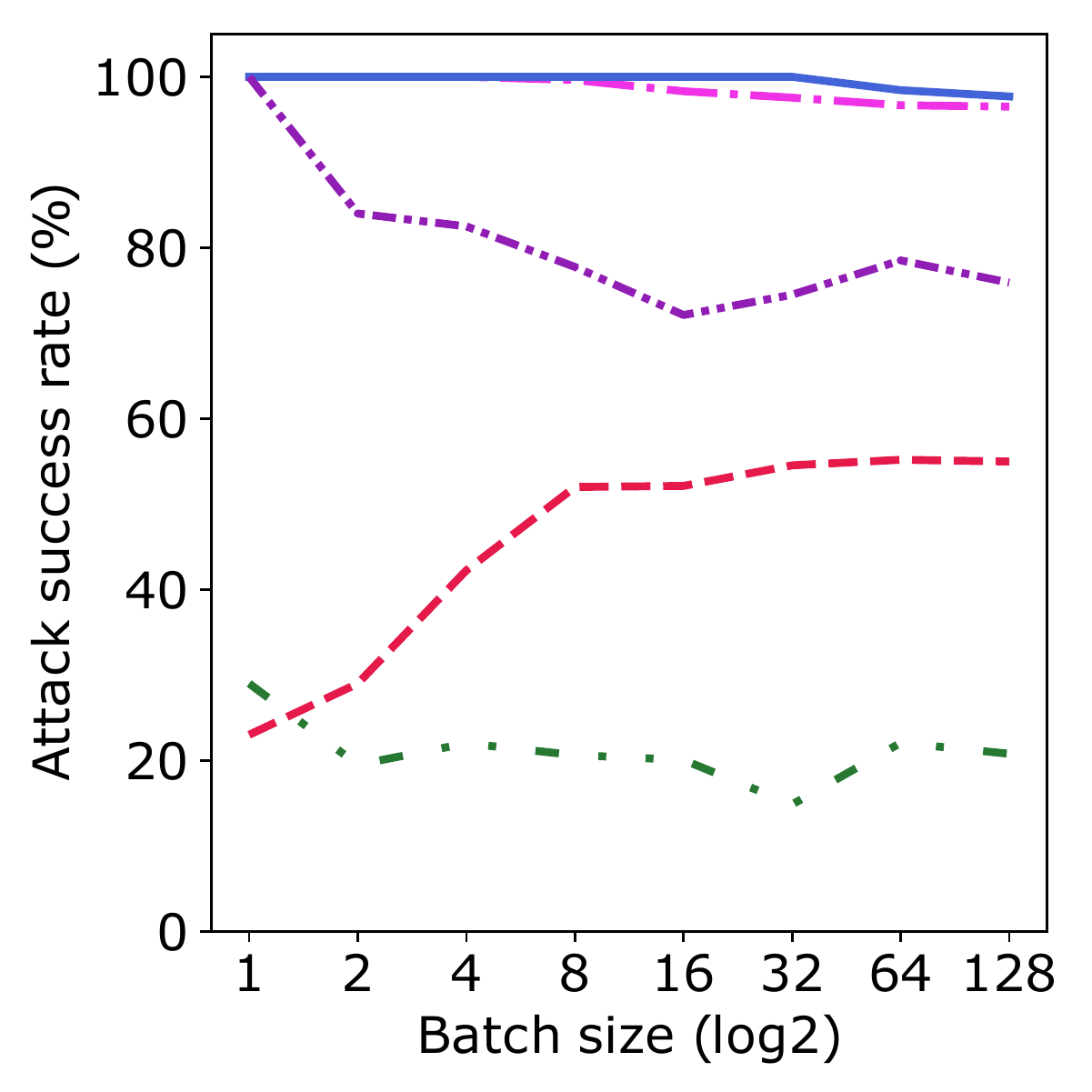} 
        \caption{CelebA - FS} 
        \label{fig:celebA_accuracy_noniid_sgd}
    \end{subfigure}

    \begin{subfigure}[b]{0.26\textwidth}
         \includegraphics[trim={0 0 0.4cm 0}, clip, width=\textwidth]{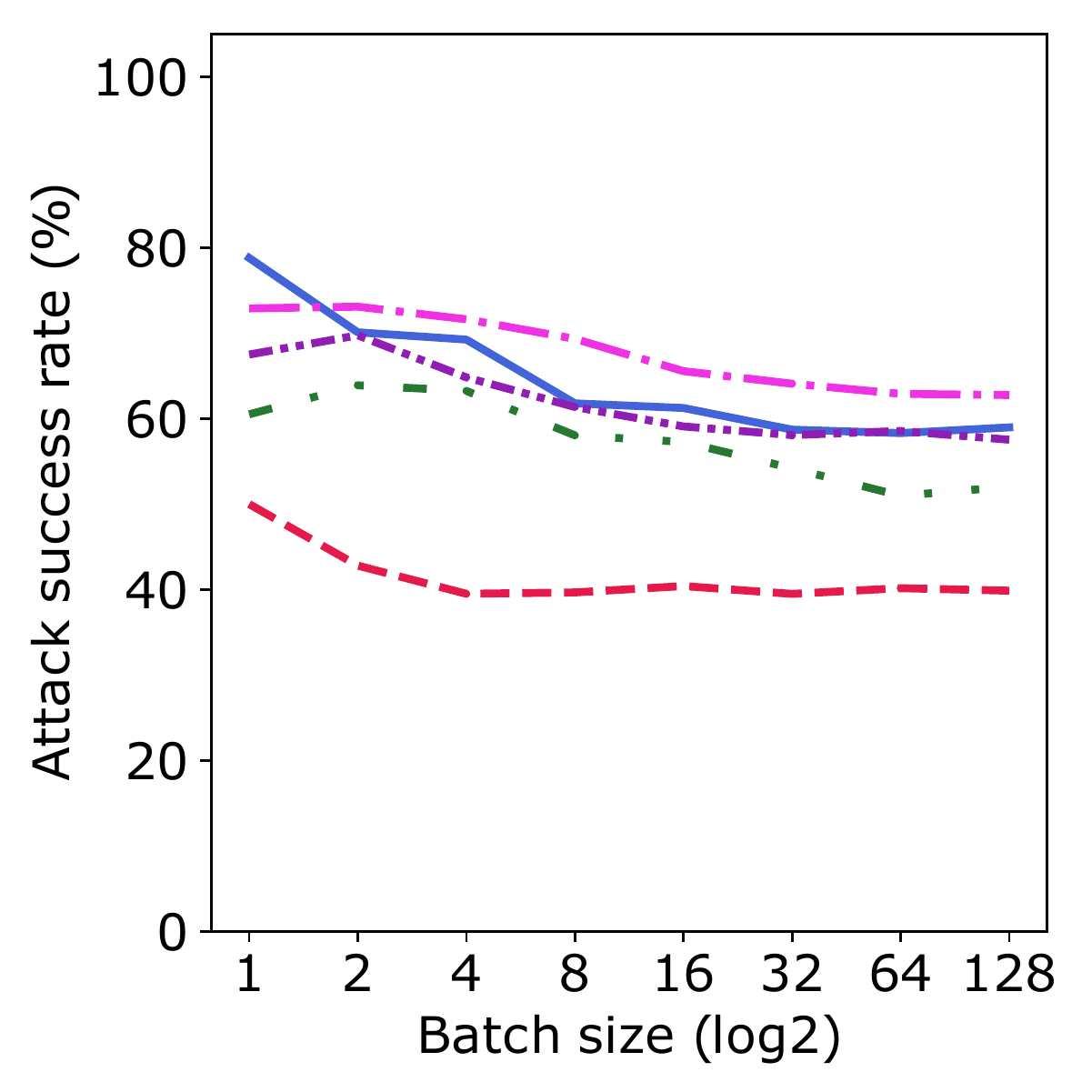} 
        \caption{MNIST - FA} 
        \label{fig:mnist_accuracy_noniid_avg}
    \end{subfigure}   
    \begin{subfigure}[b]{0.235\textwidth}
        \includegraphics[trim={1.1cm 0 0.4cm 0}, clip, width=\textwidth]{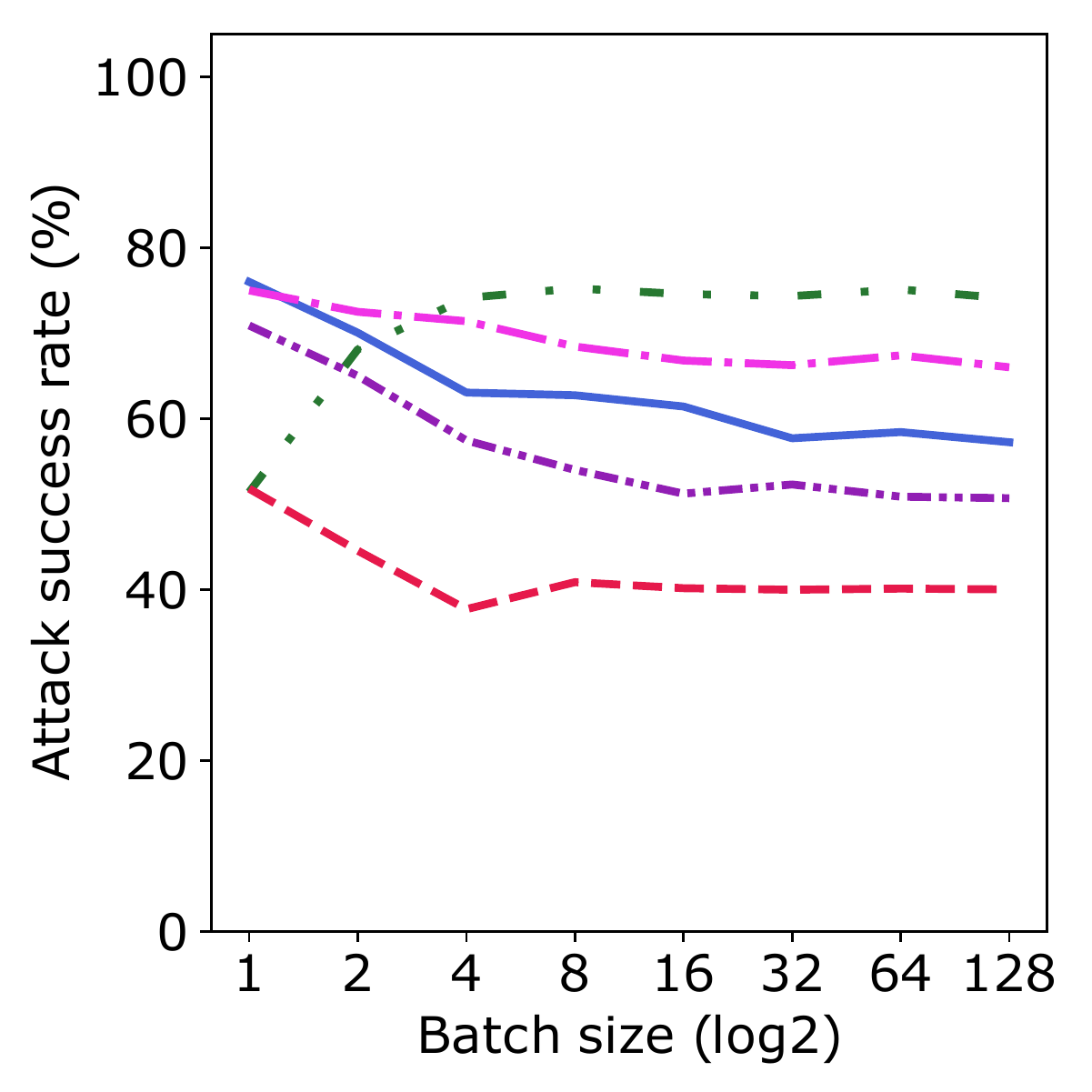} 
        \caption{SVHN - FA} 
        \label{fig:svhn_accuracy_noniid_avg}
    \end{subfigure} 
    \begin{subfigure}[b]{0.235\textwidth}
        \includegraphics[trim={1.1cm 0 0.4cm 0}, clip, width=\textwidth]{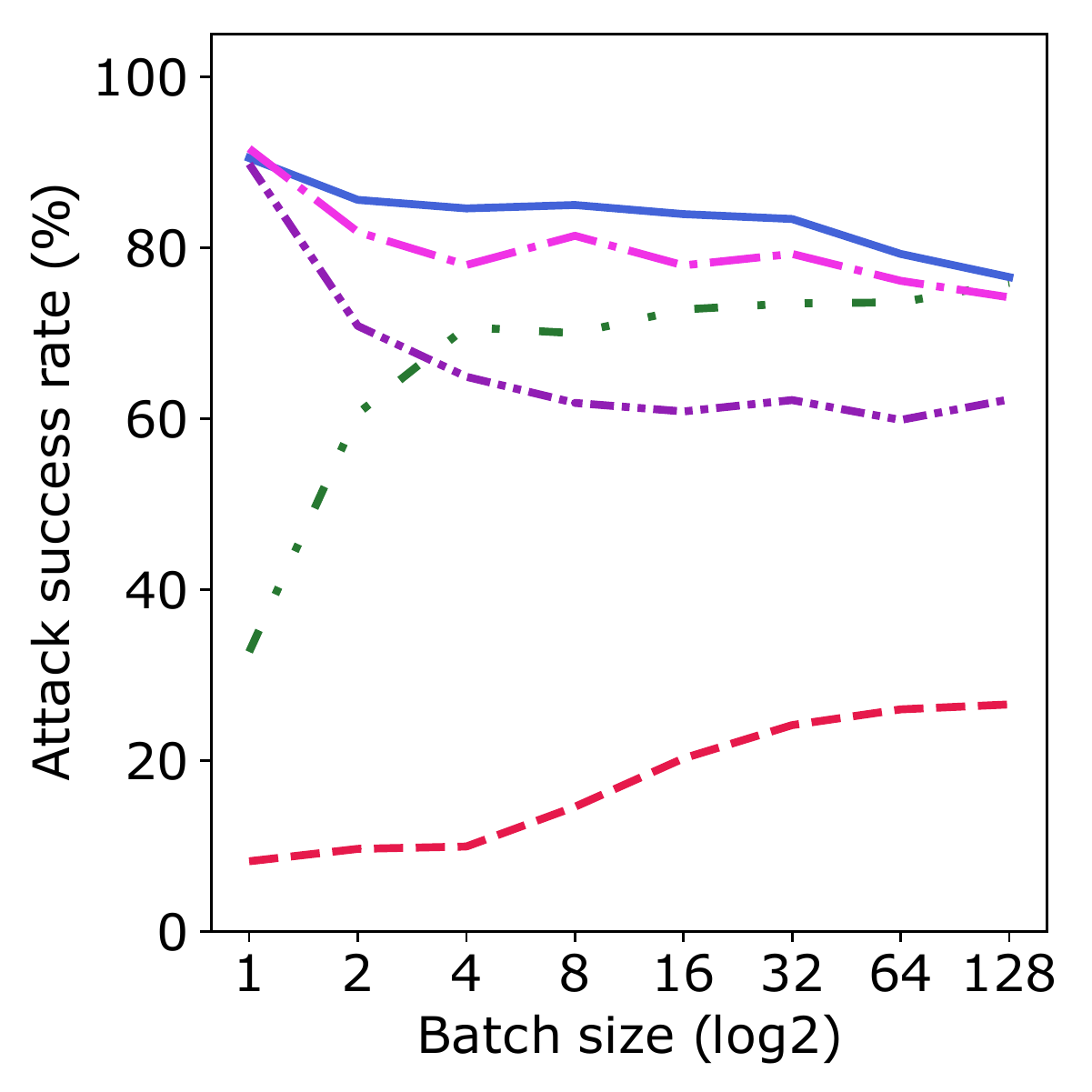} 
        \caption{CIFAR-100 - FA} 
        \label{fig:cifar_accuracy_noniid_avg}
    \end{subfigure}  
     \begin{subfigure}[b]{0.235\textwidth}
        \includegraphics[trim={1.1cm 0 0.4cm 0}, clip, width=\textwidth]{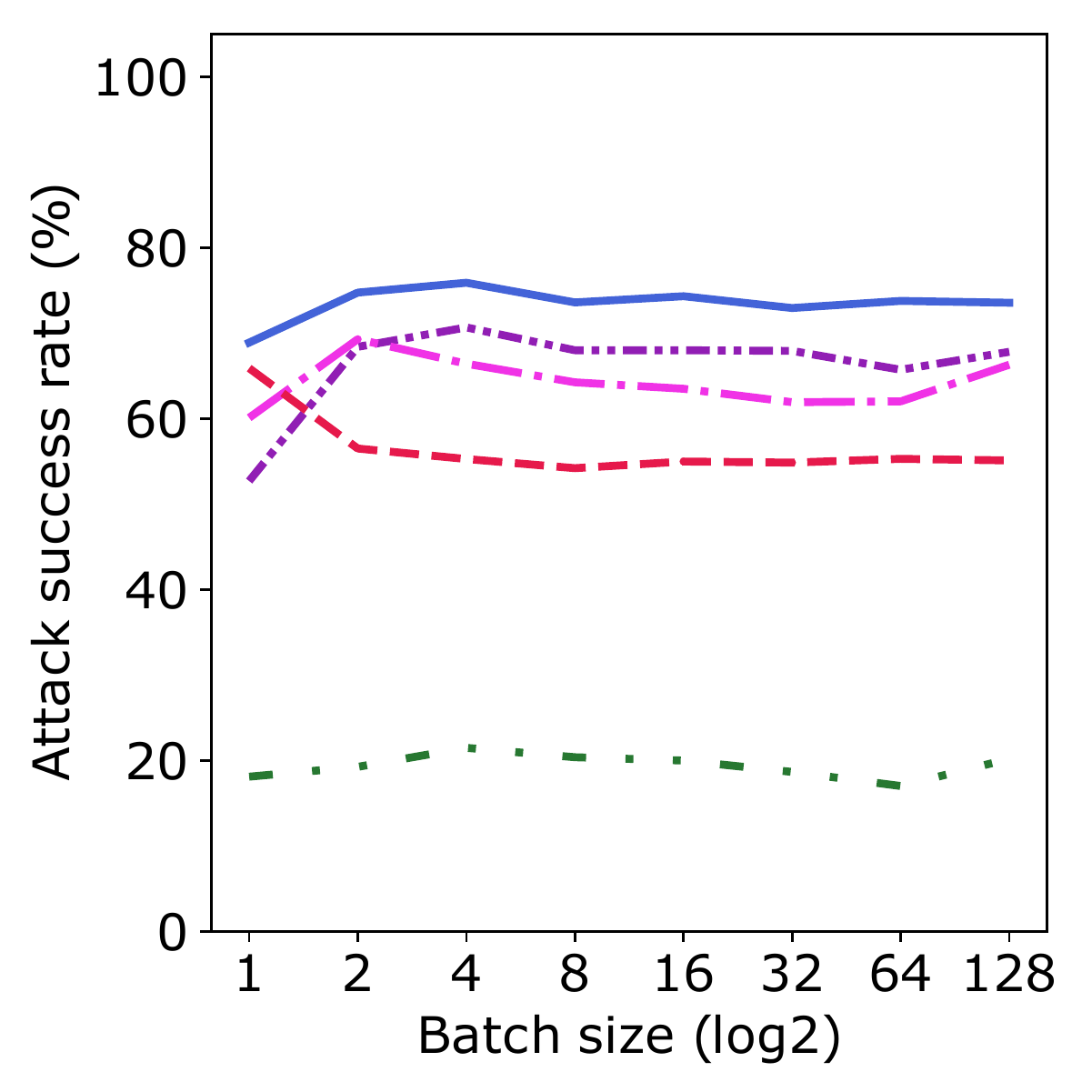} 
        \caption{CelebA - FA} 
        \label{fig:celebA_accuracy_noniid_avg}
    \end{subfigure}

    \caption{\inj{\footnotesize
    Attack success rate of (1) LLG with shared gradients, (2) LLG* with white-box model, (3) LLG+ with auxiliary knowledge, (4) DLG~\cite{zhu2019deep}, and (5) random guess on MNIST, SVHN, CIFAR-100, and CelebA. Label extraction is based on gradients generated from passing (1) one batch for FedSGD (FS) in the first row, and (2) $10$ batches for FedAvg (FA) in the second row, to a randomly initialized CNN. DLG runs for 100 iterations. LLG methods outperform the baselines in most of the cases.}}
    \label{fig:asr}
\end{figure*}

\inj{We run our experiments under two \gls{FL} algorithms, namely, FedSGD and FedAvg~\cite{mcmahan2016communication}.
For FedSGD, we pass one batch to the model and attack the generated gradients.
While for FedAvg, we feed the model with $10$ batches and attack the aggregated gradients, i.e., the sum of the gradients over $10$ iterations.}
\inj{During our experiments, we observed a very limited difference in the \gls{ASR} of the \gls{LLG} attacks for balanced and unbalanced batches.
Therefore, and because the unbalanced data is closer to real-world scenarios~\cite{mcmahan2016communication}, we focus on presenting the results of the unbalanced data case,} 
\injj{while providing a part of the balanced data results (for FedSGD) in the Appendix, Figure~\ref{fig:grad_acc_balanced}.}

\begin{figure*}[t]
    \centering  
    \begin{subfigure}[t]{0.09\textwidth}
        \includegraphics[width=\textwidth]{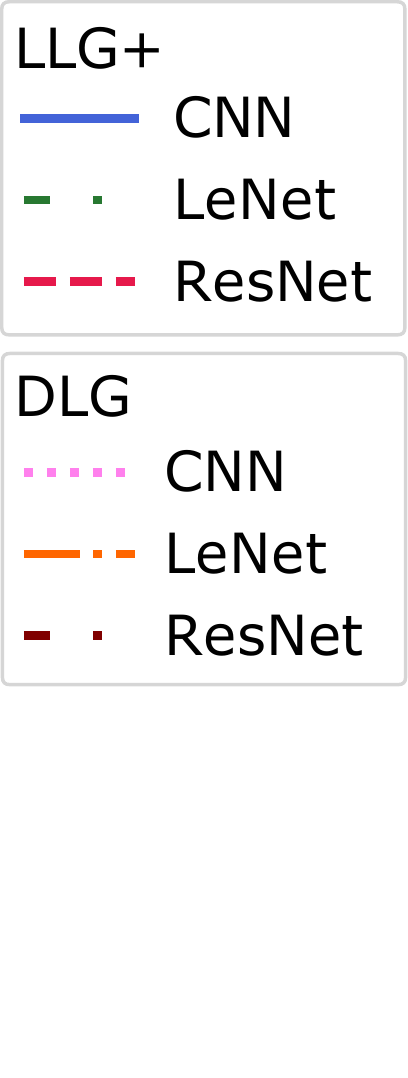}
    \end{subfigure}  
    \begin{subfigure}[t]{0.33\textwidth}
        \includegraphics[trim={0 0 0.2cm 0}, clip, width=\textwidth]{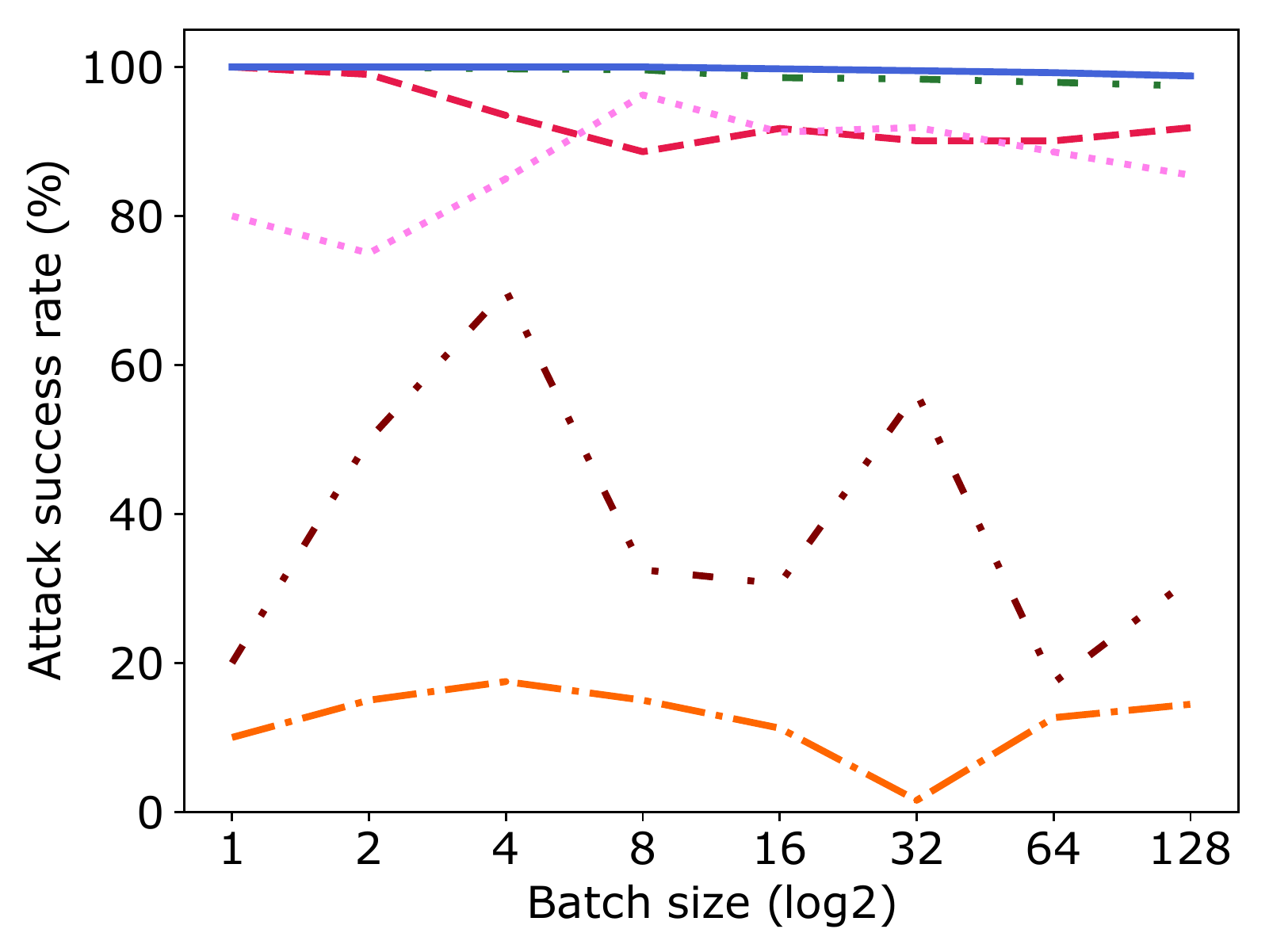}
        \caption{FS}
        \label{fig:archi_fedsgd}
    \end{subfigure}   
    \begin{subfigure}[t]{0.315\textwidth}
        \includegraphics[trim={0.9cm 0 0.2cm 0}, clip, width=\textwidth]{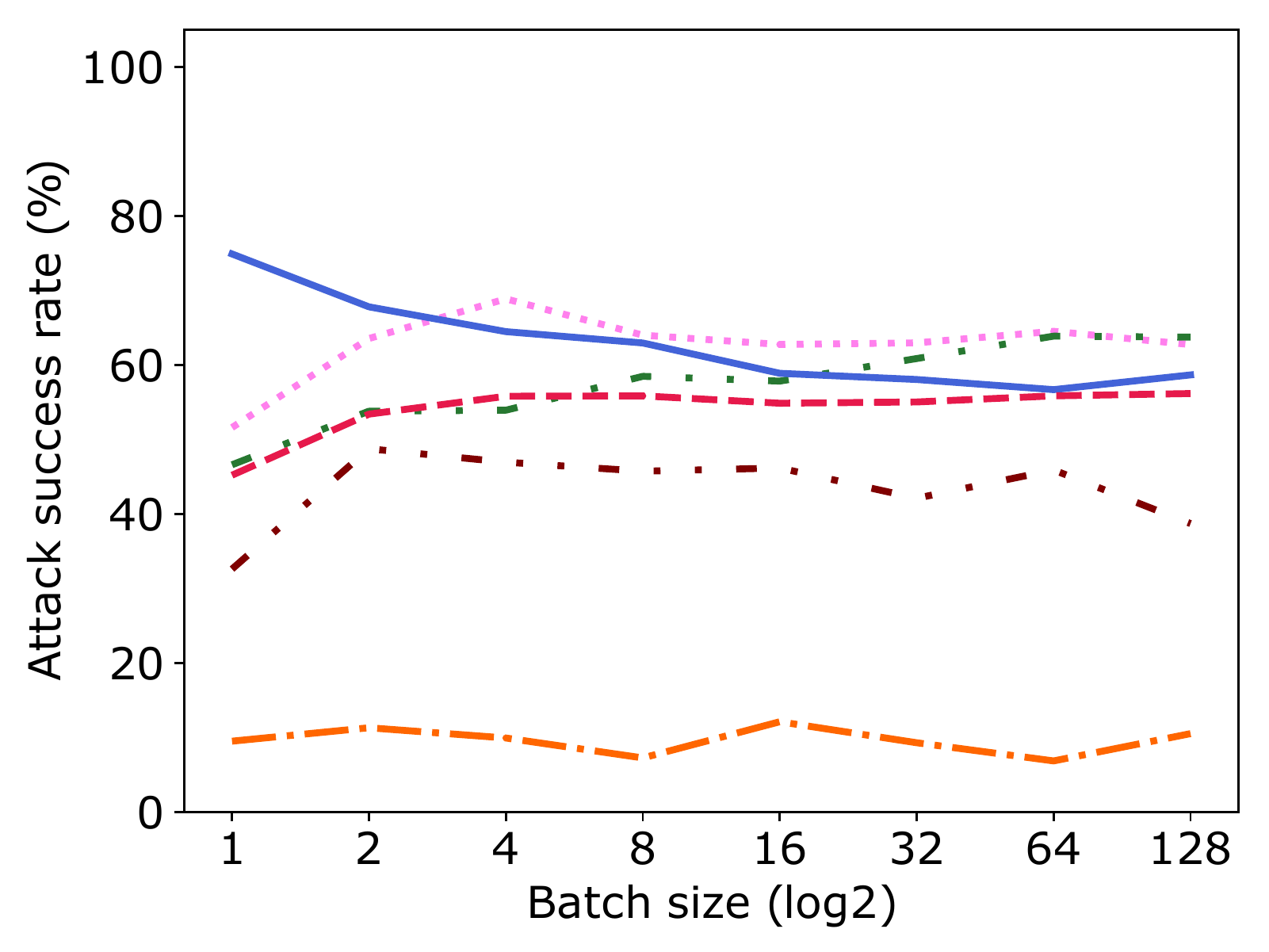}
        \caption{FA}
        \label{fig:archi_fedavg}
    \end{subfigure} 

    \caption{\footnotesize Attack success rate of \gls{LLG}+ and DLG on unbalanced batches of different sizes from MNIST with different model architectures: CNN, LeNet, ResNet20.
    \inj{For FedSGD (FS),} \gls{LLG}+ achieves around 100\% accuracy on CNN and LeNet while its accuracy slightly decreases on ResNet20. 
    \inj{DLG achieves $>80\%$ for most batch sizes on CNN, while drops remarkably on more complex architectures i.e. LeNet and ResNet20.}
    \inj{For FedAvg (FA), the ASR of \gls{LLG}+ is slightly different form architecture to another, while DLG shows higher sensitivity to the architectures.} }
    \label{fig:model_archi}
\end{figure*}

\begin{description}[leftmargin=0cm]
    \item[FedSGD.] Figures~\ref{fig:asr}~(a-d) illustrate the \gls{ASR} scores (y-axis) with batches of different sizes (x-axis).
We can see that all \gls{LLG} variants show some level of \gls{ASR} degradation \inj{when the batch size increases.
However, it appears to be stabilized to some extent for bigger batches, e.g., $64$ and $128$.
This is due to the fact that the first step of the algorithm (see Section~\ref{sub:algo}) is based on Property~1 and yields 100\% correct labels.
This step extracts a maximum of $n$ labels.
Thus, its results are dominant when $B \leq n$.
Where as if $B > n$, the second and third steps, which are based on heuristic estimations, contribute more to the final extracted labels. 
As a result, we notice a degradation of the \gls{ASR}.
However, the different batch sizes do not seem to massively affect the correctness of the results of these steps.
This might be explained by the fact that the batch size $B$ is always considered as a parameter in the heuristic estimations of the impact and offset.
}

\quad Overall, \gls{LLG}+ outperforms all the other \gls{LLG} variants and DLG.
The \gls{LLG} and \gls{LLG}* scores range from 100\% to a minimum of 77\% across the different datasets.
Whereas \gls{LLG}+ remarkably exhibits a high level of stability for various batch sizes and number of classes (in datasets) with an $ASR >98\%$.
This mainly reflects the quality of our estimation methods for impact and offset.

\quad In contrast, \gls{DLG} achieves varying accuracy scores. However, 
\inj{no clear behavior can be concluded w.r.t. the changes in the batch sizes.}
This might be due to the fact that \gls{DLG} requires a training phase, which is highly sensitive to model initialization, i.e., it might fail to converge for some randomly initialized models or might take different amounts of time for reaching a specific accuracy, unlike \gls{LLG}, which yields more deterministic results, while at the same time being orders of magnitude faster.
\inj{For example, the execution time of those experiments illustrated in Figure~\ref{fig:asr}~(a) is as follows: LLG 54s, LLG* 32.2m, LLG+ 14.6m, DLG 17.4h, and Random 50s, using a Tesla GPU V100-SXM3-32GB.
It is worth mentioning that LLG* requires more time than LLG+ due to the dummy images generation.}

\quad The \gls{ASR} of each \gls{LLG} attack is similar to some extent on MNIST and SVHN respectively.
This can be explained by the fact that both datasets have the same number of classes, i.e. 10.
On CIFAR-100 (100 classes), interestingly, we notice that \gls{LLG} performs quite closely to \gls{LLG}+ (both have $ASR >96\%$), as shown in Figure~\ref{fig:asr}~(c), \inj{while it drops to around 75\% on CelebA (5 classes).
This observation suggests that \gls{LLG} performs better for datasets with a bigger number of classes.
This can be explained by the fact that \gls{LLG} solely depends on the quality of the impact parameter, which is derived from Eq.~\eqref{eq:gi_m} under the assumption that the untrained model performs poorly.
This assumption is more valid when the number of classes is bigger, as we explained earlier in the proof of Property~2, Section~\ref{sec:gradients}.
Therefore, the estimation of the impact yields better results leading to higher \gls{ASR}.}

\quad \gls{LLG}*, with its dummy data for the parameter estimation, shows a notable drop on CIFAR-100.
It is known that the complexity of CIFAR-100 images is higher than the one of MNIST and SVHN.
Therefore, we can conclude that the complexity of the dataset might influence \gls{LLG}* in a negative way, 
while it has no observable effect on \gls{LLG} and \gls{LLG}+.
\inj{
For DLG, we notice in Figure~\ref{fig:asr}~(d) a remarkable decrease in accuracy on CelebA.
This can be due to the fact that the images are of higher dimensions ($178 \times 218$), unlike the other datasets.
Thus, the convergence of the attack is much more difficult.}

\item[FedAvg.]
\inj{
In Figures~\ref{fig:asr}~(e-h), we can see that the \glspl{ASR} of all the \gls{LLG} variants considerably decrease comparing with FedSGD, ranging between $55\%$ and $90\%$.
This is expected as the shared gradients are generated from multiple iterations ($10$ batches). 
Thus, the correlation between the gradient values and label occurrences is less prominent.
In other words, the gradients are accumulated several times over iterations, such that the correlation (Property~1 and 2) become more difficult to detect and exploit.
However, the \gls{LLG} attacks achieve higher \glspl{ASR} than the random guess on all the datasets, thus, they are still posing a serious threat.
The superiority of \gls{LLG}+ is maintained on CIFAR-100 and CelebA, while in MNIST and SVHN, \gls{LLG}* and DLG surprisingly perform the best. 
}
\end{description}

\subsection{Model architecture}

Here, we study the influence of the model architecture on the studied attacks; 
for that, we consider two models besides our default \gls{CNN}:
\begin{enumerate*}[label*=(\arabic*)]
    \item LeNet~\cite{lecun1998gradient}, a basic \gls{CNN} that contains 3 convolutional layers with 2 maximum pooling layers as shown in the Appendix, Table~\ref{tbl:lenet_arch}.
    \item ResNet20~\cite{he2016deep}, a successful residual architecture with convolutions, which introduces the concept of \enquote{identity shortcut connection} that skips one or more layers to avoid the problem of vanishing gradients in deep neural architectures.
    ResNet20 contains 20 layers in total: 9 convolutional layers, 9 batch normalization, and 2 linear layers.
\end{enumerate*}

\inj{Both aforementioned architectures, alongside their principal components, namely, convolutions and residual blocks, have achieved and contributed to state-of-the-art results on several classification tasks.
}

The two main conditions for Property~1 
to hold are: 
\begin{enumerate*}[label*=(\arabic*)]
    \item using the cross-entropy loss and
    \item having a non-negative activation function in the last layer before the output.
\end{enumerate*} 
Thus, we assume the labels extracted in the first step of the attack (see Algorithm~\ref{algo:llg}, Line 1-5) based on this property to be correct regardless of the rest of the model architecture.
While the next steps of \gls{LLG} are based on the impact, \inj{offset, and their estimations, which might be of different accuracy from one model architecture to another.} 
To run our analysis, we use MNIST with varying batch sizes and measure the \gls{ASR} of \gls{LLG}+ \inj{and DLG.}

\begin{figure*}[t]
    \centering 
        \begin{subfigure}[t]{0.1\textwidth}
        \includegraphics[width=\textwidth]{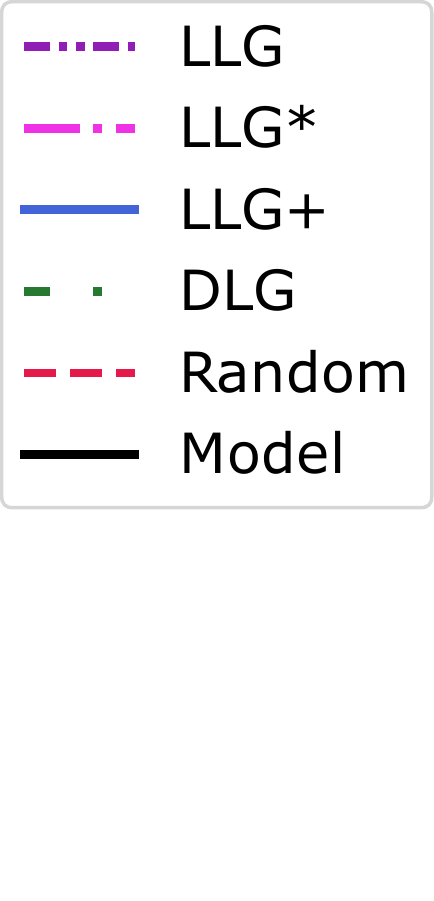}
    \end{subfigure}  
    \begin{subfigure}[t]{0.35\textwidth}
        \includegraphics[trim={0 0 0.2cm 0}, clip, width=\textwidth]{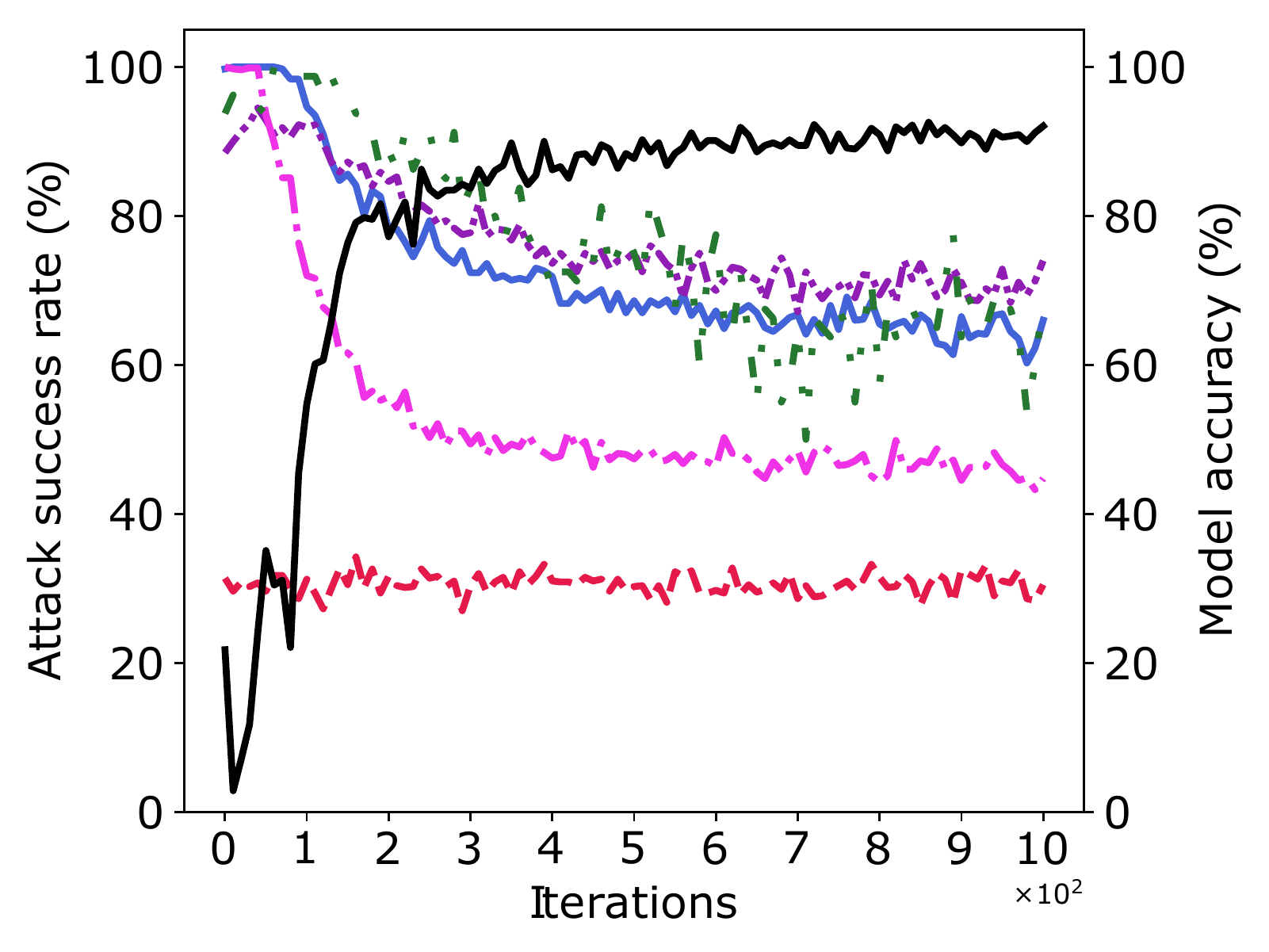}
        \caption{FS}
        \label{fig:trainig_sgd_mnist}
    \end{subfigure}  
    ~~
    \begin{subfigure}[t]{0.35\textwidth}
        \includegraphics[trim={0 0 0.2cm 0}, clip, width=\textwidth]{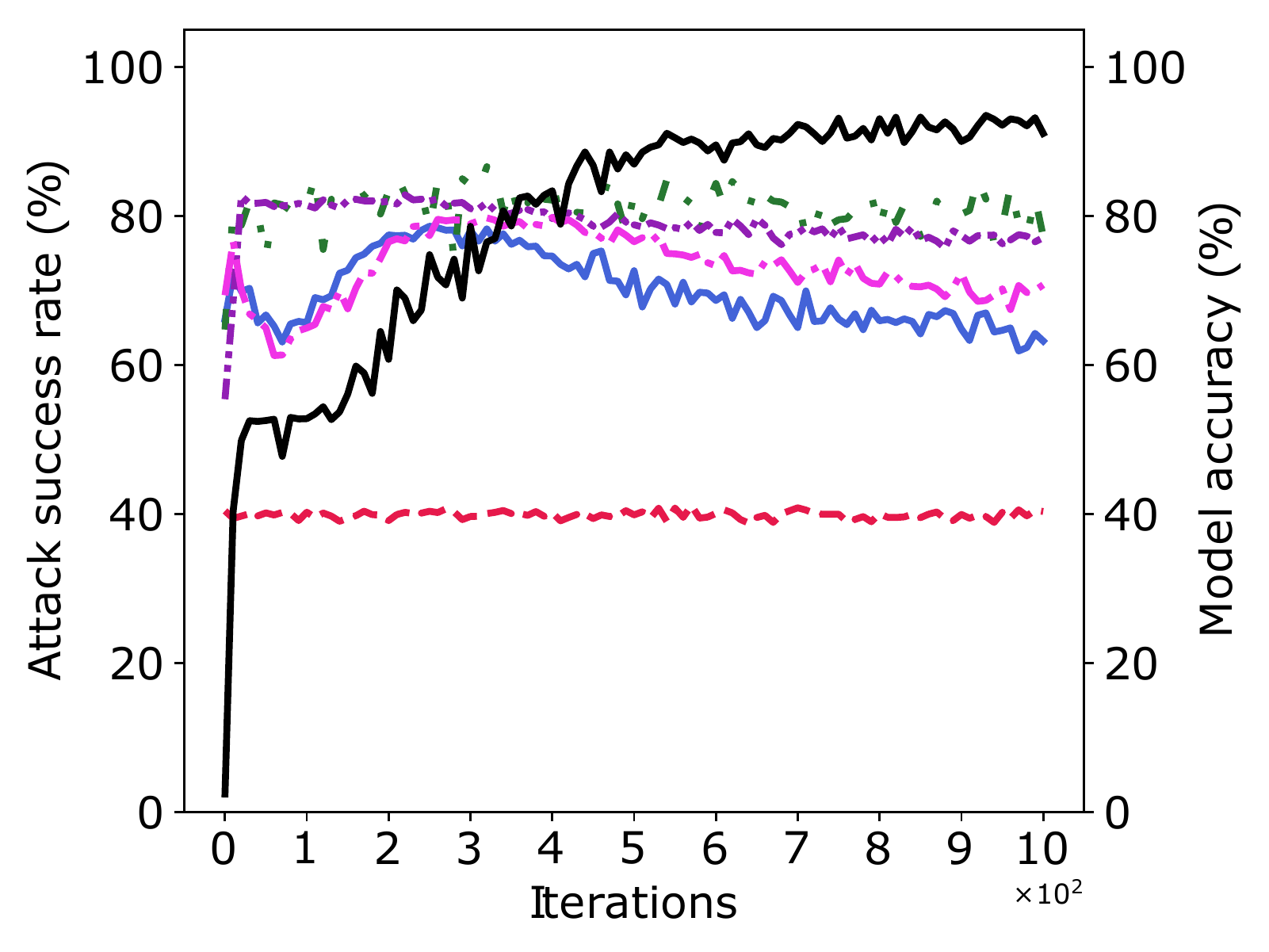}
        \caption{FA}
        \label{fig:trainig_avg_mnist}
    \end{subfigure} 

    \caption{\footnotesize Influence of model convergence status on \gls{ASR} of LLG, LLG*, LLG+, DLG, and random guess for CNN with unbalanced batches from MNIST dataset. 
    On the left y-axis it is plotted the attack success rate, while on the right y-axis, it is plotted the model test accuracy. The number of training iterations ($\times10^3$) is on the x-axis. All different LLG methods achieve remarkable success rates even if the models are well-trained and gradients become less informative.}
    \label{fig:grad_acc_trained}
\end{figure*}

\begin{description}[leftmargin=0cm]
    \item[FedSGD.] 
    As we can see in Figure~\ref{fig:model_archi}~(a), \gls{LLG}+ performs best on \gls{CNN} and LeNet, achieving approximately 100\% of success rate, while a degradation starts from batches with size $>2$ for ResNet20. 
    This is mainly due to the \inj{residual blocks} in the ResNet20 architecture that prevents the vanishing gradients problem in deep neural networks. 
    In other words, ResNet20 implicitly alters and controls the \inj{range of the} gradient values in order to not let them vanish (gradients close to zero) or to explode (gradients \inj{go towards} $+\infty$) during training. 
    \inj{
    This manipulates our definitions of the impact and offset parameters in Eq.~\eqref{eq:impact_def} and~\eqref{eq:offset_def}, thus, directly affecting the attack performance.} 
    
    \quad \inj{On the other hand, DLG shows much higher sensitivity towards the model architecture.
    As we can see it achieves $>80\%$ for most batch sizes on CNN, while dropping remarkably on more complex models, LeNet and ResNet20. 
    Such a strong influence of the model architecture on DLG is expected, as DLG includes an optimization phase, where optimizing complex models typically requires much more iterations.}
    
    \item[FedAvg.] 
    \inj{
    In Figure~\ref{fig:model_archi}~(b), we observe that under small batch sizes $B \leq 16$, the \gls{ASR} of \gls{LLG}+ is higher for CNN.
    While for bigger batches, \gls{LLG}+ only slightly differs from one model to another.
    This supports the finding that the model architecture has limited effect on \gls{LLG}+.
    In contrast, DLG shows again higher sensitivity with bigger variance of the \gls{ASR} over the different architectures.
    }
\end{description}

\subsection{Model convergence status}\label{subsec:status}

The gradients guide the model towards a local minimum of the loss function.
As the model converges to this minimum, the information included in the gradients becomes less prominent.
Therefore, we expect the convergence status of the model to have a strong influence on the attack effectiveness. 
All the previous experiments are conducted \inj{in one communication round,
i.e., the gradients are generated and shared with the server only once.}
In this section, we go further with training the model and observe the implications on the attack.

\inj{We train the model in a federated setting, where the data of MNIST is distributed among $750$ users, each has $80$ unbalanced data samples.
The server selects randomly $100$ users for every communication round to train the global model locally and share their gradients.}
The CNN model is trained with batches of size $8$ for $10^3$ iterations.
\inj{We chose the batch size $8$ to be able to apply the DLG attack in its most effective setting $B \leq 8$~\cite{zhu2019deep}.
In every communication round, we attack the shared gradients of one target user (victim) with DLG and \gls{LLG} variants, where the impact and offset are estimated dynamically.}
Figures~\ref{fig:grad_acc_trained}~(a,b) depict the attacks \gls{ASR} (on the left y-axis) versus the model accuracy at testing time (on the right y-axis), while the x-axis represents the number of training iterations.

\begin{figure*}[t]
    \centering  
    \begin{subfigure}[b]{\textwidth}
    \includegraphics[width=0.987\textwidth]{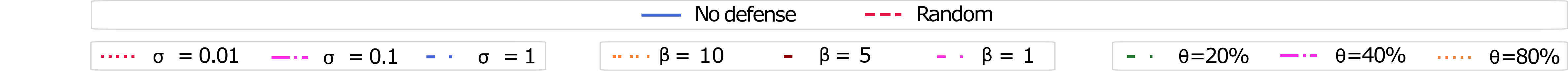}
    \end{subfigure}  
    
    \begin{subfigure}[b]{0.337\textwidth}
        \includegraphics[trim={0 0 0.2cm 0}, clip, width=\textwidth]{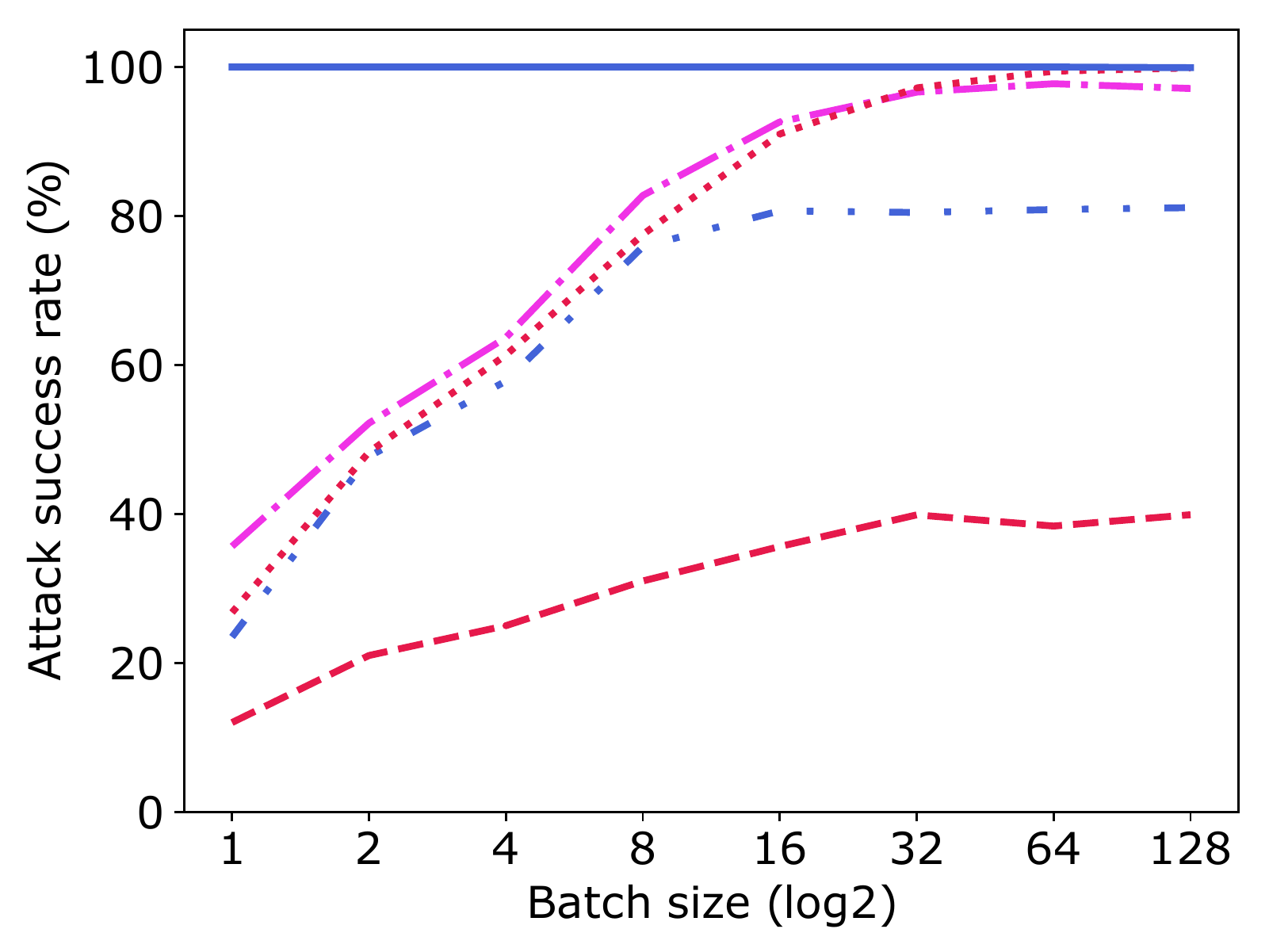}
        \caption{Pure noise - FS}
        \label{fig:sgd_noise_gaussian_unbalanced}
    \end{subfigure}   
    \begin{subfigure}[b]{0.318\textwidth}
        \includegraphics[trim={0.9cm 0 0.2cm 0}, clip, width=\textwidth]{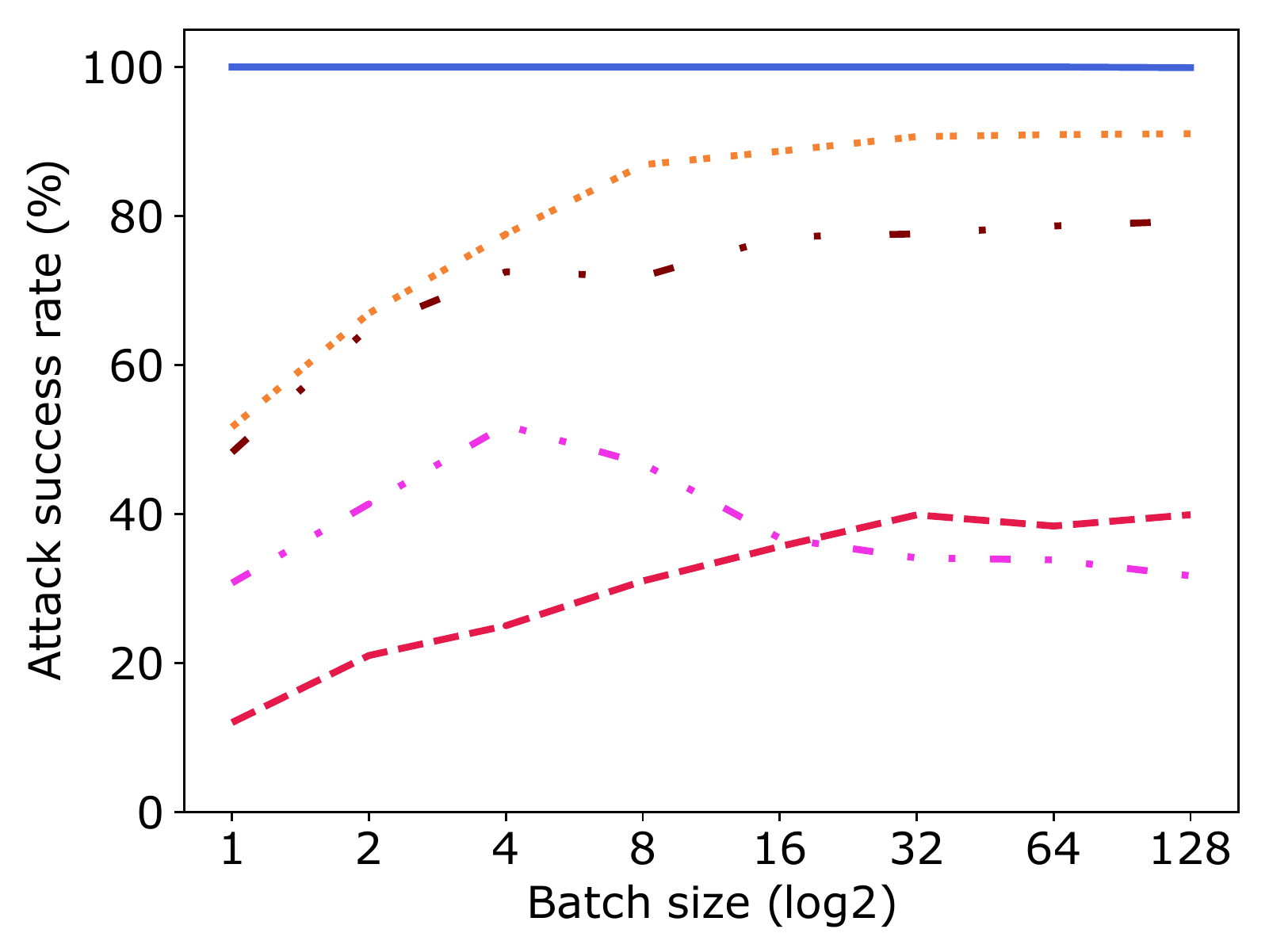}
        \caption{Differential privacy - FS}
        \label{fig:sgd_dp_unbalanced}
    \end{subfigure} 
    \begin{subfigure}[b]{0.318\textwidth}
        \includegraphics[trim={1cm 0 0.2cm 0}, clip, width=\textwidth]{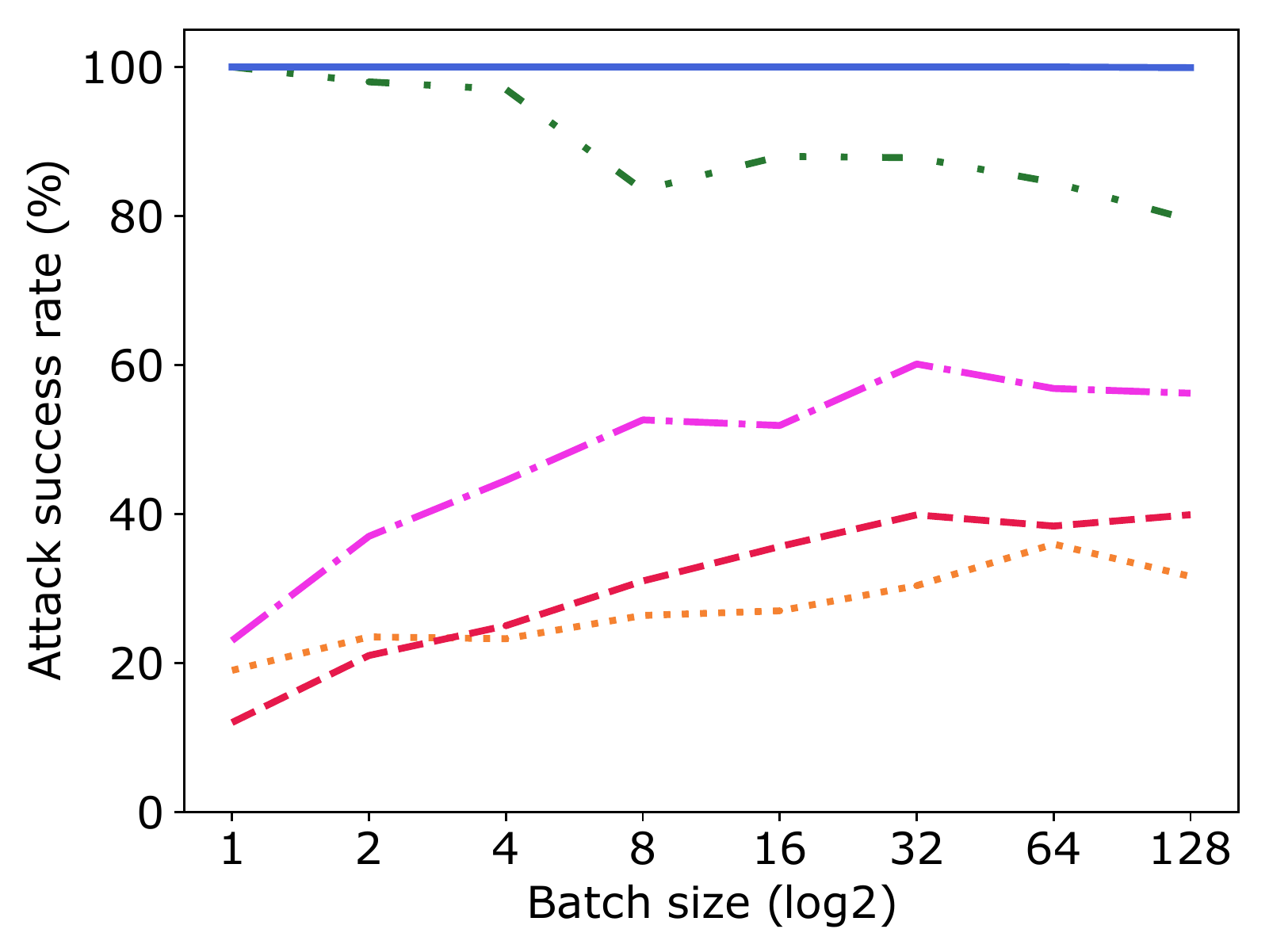}
        \caption{Compression - FS}
        \label{fig:sgd_compress_unbalanced}
    \end{subfigure} 

    \begin{subfigure}[b]{0.337\textwidth}
        \includegraphics[trim={0 0 0.2cm 0}, clip, width=\textwidth]{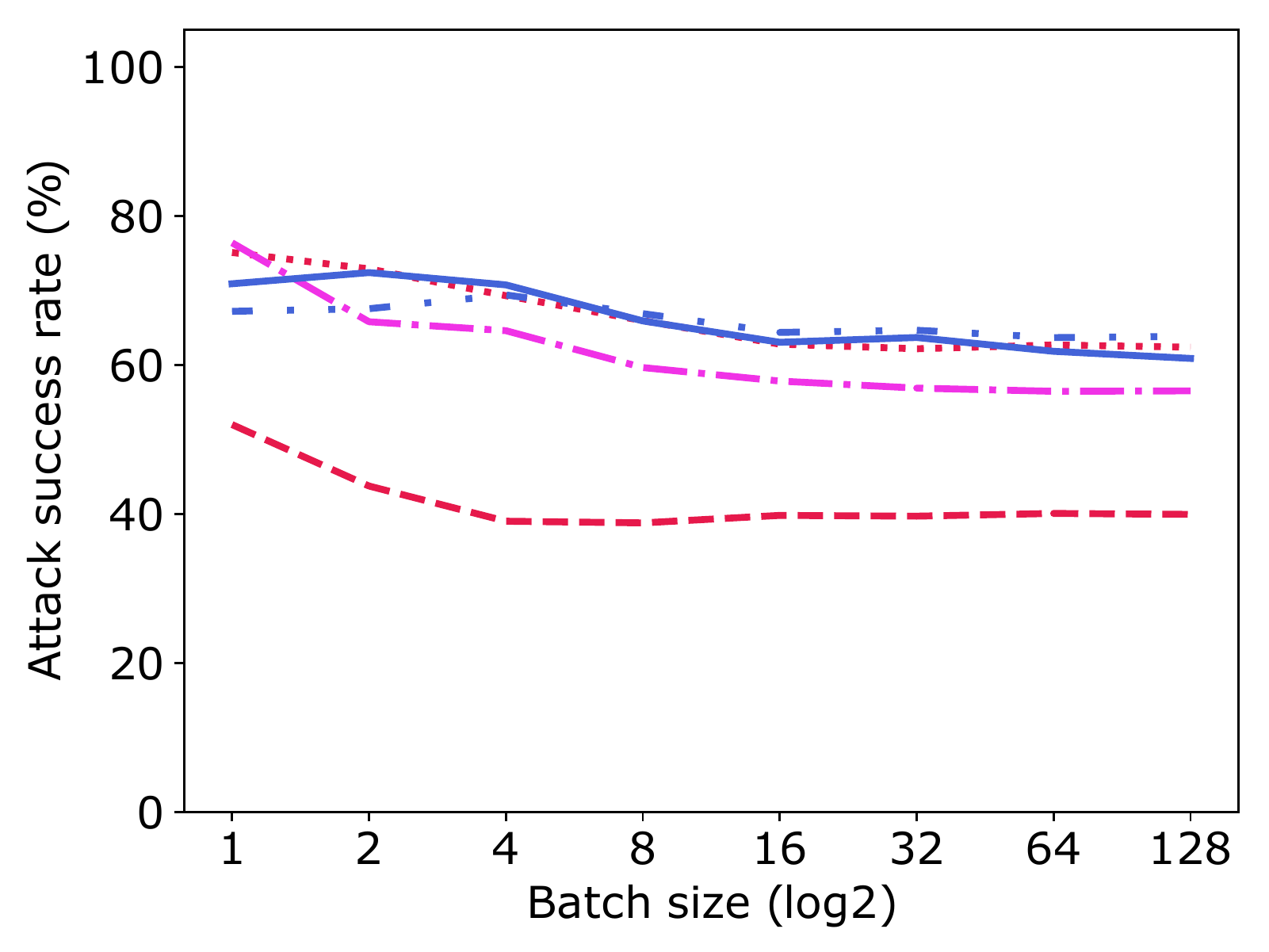}
        \caption{Pure noise - FA}
        \label{fig:fedavg_noise_gaussian_unbalanced}
    \end{subfigure}   
    \begin{subfigure}[b]{0.318\textwidth}
        \includegraphics[trim={0.9cm 0 0.2cm 0}, clip, width=\textwidth]{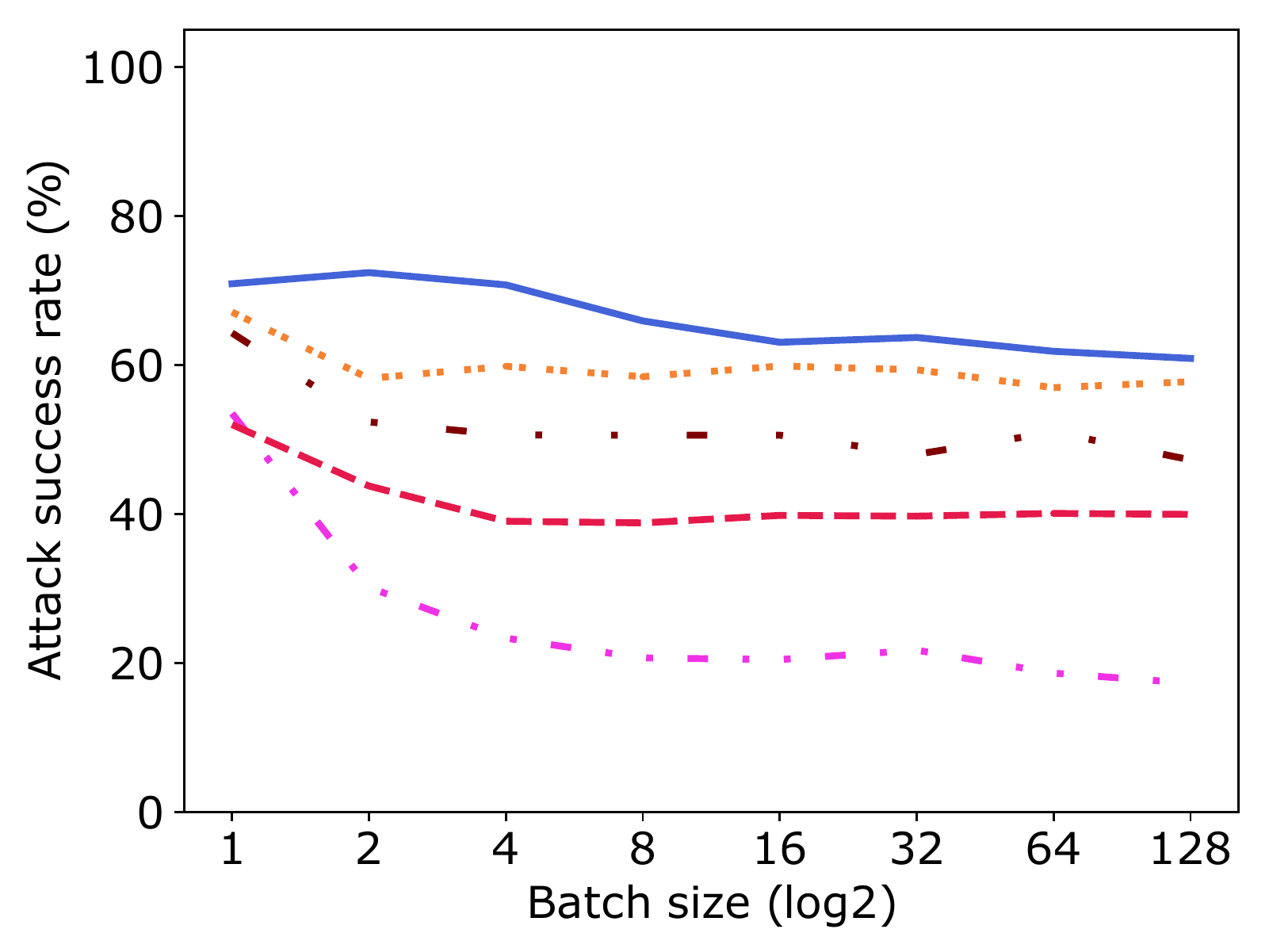}
        \caption{Differential privacy - FA}
        \label{fig:fedavg_dp_unbalanced}
    \end{subfigure} 
    \begin{subfigure}[b]{0.318\textwidth}
        \includegraphics[trim={1cm 0 0.2cm 0}, clip, width=\textwidth]{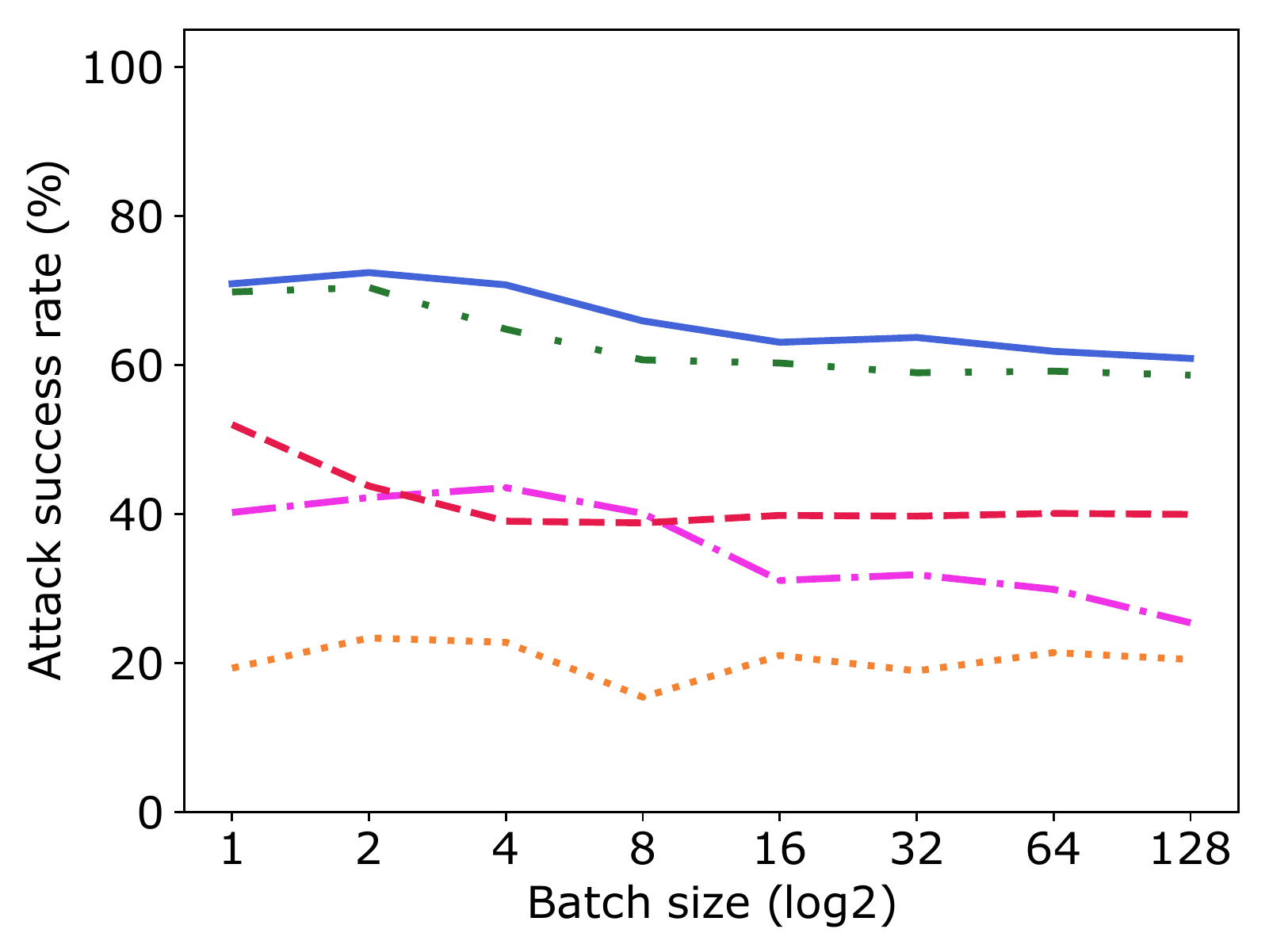}
        \caption{Compression - FA}
        \label{fig:fedavg_compress_unbalanced}
    \end{subfigure} 
    
    \caption{\footnotesize Effectiveness of different defenses against \gls{LLG}+ on an unbalanced batch from MNIST with a randomly initialized CNN: (a)~defense by adding Gaussian distributed noise to gradients with $\sigma \in \{0.01, 0.1, 1\}$, (b)~defense by \injj{user-side differential privacy with $\sigma = 0.1$,} and clipping bound $\beta \in \{1, 5, 10\}$, (c)~defense by pruning gradients with varying compression ratios $\theta \in \{20\%, 40\%, 80\%\}$. Pure noise is not successful in eliminating the risk completely, since \gls{LLG}+ maintains a higher \gls{ASR} than the random guess.
    \inj{While, differential privacy mitigates the attack with $\beta=1$ for FedAvg (FA), and for FedSGD (FS) when batch size $B \geq 16$.
    Gradient compression is effective in FedSGD when a high compression ratio ($ \ge 80\%$) is used with $B \geq 4$.
    For FedAvg, even the compression ratio $40\%$ with $B \geq 8$ is an effective defense.}}
    \label{fig:defenses}
\end{figure*}

\begin{description}[leftmargin=0cm]
   \item[FedSGD.]
   We can see in Figure~\ref{fig:grad_acc_trained}~(a) that the growth of the model accuracy incurs a notable decrease of the \gls{ASR} for all \gls{LLG}s. 
    However, although the model converges close to 93\% accuracy, \gls{LLG} and \gls{LLG}+ keep achieving $\gls{ASR}>60\%$, considerably higher than the random guess which is around 32\%.
    Meaning, the attacks are still able to take advantage of the reduced information in gradients over the course of the whole training process. 
    \inj{Similarly, DLG shows degradation in accuracy, yet it remains effective for well-trained models.} 
   \item[FedAvg.]
   \inj{
    Figure~\ref{fig:grad_acc_trained}~(b) shows more stability of \gls{ASR} over the training process, especially for \gls{LLG} and DLG.
    Even in the early stages of the training, the multiple local iterations in FedAvg improve the model accuracy, thus, make the accumulative gradients less informative.
    Therefore, the attacks start with lower \gls{ASR} comparing with FedSGD.
    However, this leads also to mitigating the notable degradation of \gls{ASR} observed in Figure~\ref{fig:grad_acc_trained}~(a).
    \gls{LLG}* and \gls{LLG}+ exhibit volatile behavior in the early iterations, where they have an increasing success rate between iteration $100$ and $300$.
    Then, they decrease again from $80\%$ to close to $70\%$ and $60\%$, respectively.  
    \injj{Interestingly, DLG maintains a high success rate (around $80\%$) outperforming the \gls{LLG} variants in most parts of the training process.
    This shows that DLG is less sensitive to convergence status under FedAvg and thus can cope with the decreasing amount of information in the gradients.}
    Overall, all the attacks stay effective with $\gls{ASR}>40\%$, which is the random guess success rate.
    }
\end{description}

\subsection{Defense mechanisms}\label{sec:defense}

As \gls{LLG} is mainly based on the gradients, thus, sensitive to changes in their values, obfuscating them can be a direct mitigation mechanism.
In this section, we use two obfuscation techniques: noisy gradients and gradients compression.
We apply these techniques on the user side before sharing the gradients with the server and thus, protecting against external eavesdroppers and curious servers.
Then, we attack the gradients of one target user in one communication round for a randomly initialized \gls{CNN} model (untrained).
In general, applying obfuscation techniques incurs a loss in the model accuracy.
To cover this aspect, we train the model to convergence under conditions similar to those in Section~\ref{subsec:status} while applying the defenses, and report its accuracy.

\subsubsection{Noisy gradients}
    Many researchers consider adding noise to gradients as the de facto standard for privacy-preserving \gls{ML}~\cite{li2019vision}.
    \inj{In this experiment, we evaluate \gls{LLG}+ against two techniques of  noise addition:
    \begin{enumerate*}[label*=(\arabic*)]
        \item Pure noise: we add noise on gradients before sharing, similar to~\cite{Wei2020,zhu2019deep}, where no formal privacy properties are guaranteed, and
        \item differential privacy: following  
        \injj{differentially private \gls{FL}~\cite{geyer2017differentially},}
        we clip the gradients to bound their sensitivity, then, we add noise to them.
        The clipping is defined as 
        $\nabla \bm{W} \leftarrow \nabla \bm{W} / max\left(1, \frac{\|\nabla \bm{W}\|_2}{\beta}\right)$, where $\beta$ is the gradient norm bound.
    \end{enumerate*}}
    In both noise addition techniques, we use the Gaussian noise distribution. 
    For pure noise, the standard deviation of the noise distribution is $\sigma \in \{0.01, 0.1, 1\}$ with central~0.
    For differential privacy, we use $\sigma = 0.1$ and varying norm bound $\beta \in \{1, 5, 10\}$. 
    We track the privacy loss for the model trained with differential privacy using the moments accountant~\cite{Abadi2016}.
    For $100$ communication rounds and $\delta = 10^{-5}$, the privacy budget is estimated $\epsilon \approx 11.5$.

     \begin{description}[leftmargin=0cm]
       \item[FedSGD.]
    In Figure~\ref{fig:defenses}~(a), we can see that the higher the magnitude of the noise the less accurate the attack.
    This is expected as the attack partially uses the magnitude of the gradients to infer the labels following Property~2. 
    Interestingly, we observe that the noise has less effect on the attack when the batch size is increasing.
    \inj{We investigated this observation further by inspecting the values of the gradients before and after noise addition.
    Our empirical analysis showed earlier in Figure~\ref{fig:grad_dist}~(a, b) that the majority of gradients $g_i$ have values close to zero when they correspond to labels not present in the batch. 
    Adding noise to such small gradient values might lead to flipping their sign, and consequently, disrupting Property~1, which is one of the basis of the attack.
    For batch sizes $B < n$ with $n$ as the number of classes, not all the labels will be present in the batch, so the flipping effect can be prominent on the attack success rate.
    Whereas, in bigger batches, it is more likely to have more differing labels, thus, their gradients values are not close to zero.
    As a result, adding a small amount of noise does not lead to sign flipping.
    This also explains the stability of \gls{ASR} values when $B \geq n$.}
    Overall, adding noise does not eliminate the risk completely \injj{while reducing the model accuracy (see Table~\ref{tbl:acc}).} 
    As we can see, \gls{LLG}+ maintains higher \glspl{ASR} than the random guess for all the test noise scales.
    
    \quad \inj{Figure~\ref{fig:defenses}~(b) shows that adding noise of $\sigma = 0.1$ with clipping bound $\beta=1$, 
    is an effective defense against \gls{LLG}+ for batch sizes $B>16$, where the \gls{ASR} drops beyond the random guess.
    \injj{However, this leads to a significant drop in the model accuracy as shown in Table~\ref{tbl:acc}.}}
    
   \item[FedAvg.]
    \inj{Unlike in FedSGD, the magnitude of the pure noise does not have a clear effect on \gls{ASR} for FedAvg as shown in Figure~\ref{fig:defenses}~(d).
    That is due to the fact that the shared gradients are generated from $10$ batches.
    Thus, the gradient values reflect $10 \times B$ labels, which is always greater or equal $n$ for MNIST, where $n=10$. 
    Therefore, it is likely that most of the labels appear in one of the batches at least, consequently, no gradient values will be close to zero.
    As a result, the pure noise does not impact \gls{ASR} remarkably, and \gls{LLG}+ remains effective.
    In Figure~\ref{fig:defenses}~(e), we notice that noise $\sigma = 0.1$ with bound of $\beta=1$ is able to mitigate \gls{LLG}+, reducing its success rate close to $20\%$ for bigger batch sizes.
    \injj{However, the model accuracy degrades remarkably to $52.5\%$.
    Additionally, other differential privacy approaches, e.g., DP-SGD~\cite{Abadi2016} can also be applied and investigated as a defense.}}
\end{description}

\subsubsection{Gradient compression}
Gradient compression~\cite{lin2017deep,tsuzuku2018variance} prunes shared gradients with small magnitudes to zero.
Pruning some gradients reduces the information that the attack exploits to extract the labels.
In this set of experiments, we evaluate \gls{LLG}+ under various gradient compression ratios ${\theta \in \{20\%, 40\%, 80\% \}}$, i.e., $\theta$ denotes the percentage of the gradients to be discarded in each communication round with the server. 
We use the sparsification approach proposed in~\cite{lin2017deep}, where users send only the prominent gradients, i.e., with a magnitude larger than a specific threshold.
The threshold is calculated dynamically based on the desired compression ratio. 
The small gradients are accumulated across multiple communication rounds and sent only when they are large enough.

\begin{description}[leftmargin=0cm]
    \item[FedSGD.]
    Figure~\ref{fig:defenses}~(c) illustrates that when the ratio is $\le 20\%$, there is only a slight effect on the success rate of the attack.
    When the compression ratio is $80\%$, we notice that \gls{LLG}+ becomes completely ineffective for $B \geq 4$, dropping below the random guess.
    Notably, the model accuracy is maintained high $91.9\%$ in this case.
    Consequently, gradient compression with $\theta>80\%$ can practically defend against the attack while producing accurate models.
    
    \item[FedAvg.]
    Similar to FedSGD, we observe a limited effect of the ratio $\theta \leq 20\%$ in Figure~\ref{fig:defenses}~(f), whereas the ratio of $\theta = 40\%$ with $B\geq8$ can mitigate the risk of \gls{LLG}+, as well as $\theta\geq80\%$ for any batch size.
    Under both compression ratios, the model converges at high accuracy scores, $91.6\%$ and $89.3\%$, respectively.
    Additional improvements on the accuracy can be achieved by applying error compensation techniques, such as momentum correction and local gradient clipping, which are proposed in~\cite{lin2017deep}.
\end{description} 

In addition to the aforementioned defenses, cryptography-based approaches exist~\cite{bonawitz2016practical,aono2017privacy,hao2019efficient,zhu2020privacy}, which can protect gradients from external eavesdroppers and even curious servers.
However, besides the computation and communication overhead introduced by these approaches, they prevent the server from evaluating the utility and benignity of users' updates.

\begin{table}[t]
    \caption{Model accuracy while applying the defense mechanisms. 
    PN: pure noise, DP: differential privacy, GC: gradient compression.}
    \label{tbl:acc}
    \setlength\tabcolsep{6pt}
    \begin{tabular}{l|lll|lll} 
        \hline
         & \multicolumn{3}{l|}{FedSGD (Acc. = 93.3\%)} & \multicolumn{3}{l}{FedAvg (Acc. = 94.5\%)} \\
         \hline
        \textbf{PN} ($\sigma$) & 0.01 & 0.1 & 1    & 0.01 & 0.1 & 1 \\
        Acc. (\%) & 93.4 & 89.9 & $\leq$ 10.1     & 94.6 & 91.4 & $\leq$ 13.5 \\ 
        \hline
        \textbf{DP} ($\beta$) & 10 & 5 & 1     & 10 & 5 & 1\\ 
        Acc. (\%) &  89 & 86.1 & $\leq$ 52.4         & 91.2 & 90.5 & 52.5 \\
        \hline
        \textbf{GC} ($\theta\%$) & 20 & 40 & 80     & 20 & 40 & 80 \\
        Acc. (\%) & 93.4 & 93.7 & 91.9       & 92.8 & 91.6 & 89.3\\
        \hline
    \end{tabular}
\end{table}

%% file: sections/8_conclusion.tex
\section{Conclusion}\label{sec:conclusion}
We identified and formalized two properties of gradients of the last layer in deep neural network models trained with cross-entropy loss for a classification task.
These properties reveal a correlation between gradients and label occurrences in the training batch.
We investigate \glsfirst{LLG}, a novel attack that exploits this correlation and extracts the ground-truth labels from shared gradients in the FedSGD and FedAvg algorithms.
We demonstrated the validity of \gls{LLG} through mathematical proofs and empirical analysis.
Results demonstrate the scalability of \gls{LLG} to arbitrary batch sizes and number of classes.
Moreover, we showed the success rate of \gls{LLG} on various model architectures and in different stages of training.
The effectiveness of noisy gradients and gradient compression as defenses was also investigated.
Findings suggest gradient compression to be an efficient technique to prevent the attack while maintaining the model accuracy.
With this work, we hope to raise the awareness of the privacy risks associated with gradients sharing schemes, encouraging the community and service providers to give careful consideration to security and privacy measures in this context.
As future work, we are developing improvements for the attack \inj{under FedAvg and against trained models.
Additionally, we are investigating the implications of combining \gls{LLG} with DLG on the overall accuracy of the data reconstruction.}

%% file: sections/appendix.tex
\section*{Appendix}

\begin{table}[ht]
    \caption{Architecture of CNN, the default model in the experimental setting.}
    \label{tbl:cnn_arch}
    \setlength\tabcolsep{6pt}
    \begin{tabular}{lll} 
        \hline
        Layer & Size & Activation function \\
        \hline
        (input) & - & - \\ 
        Conv 2D & channels x 12 & Sigmoid \\ 
        Conv 2D & 12 x 12 & Sigmoid \\ 
        Conv 2D & 12 x 12 & Sigmoid \\ 
        \hline
    \end{tabular}
\end{table}

\begin{table}[ht]
    \caption{Architecture of LeNet network, a very common architecture adopted in computer vision.}
    \label{tbl:lenet_arch}
    \setlength\tabcolsep{6pt}
    \begin{tabular}{lll} 
        \hline
        Layer & Size & Activation function \\
        \hline
        (input) & - & - \\ 
        Conv 2D & 1 x 6 & ReLU \\
        Maxpool & 2 x 2 & - \\
        Conv 2D & 6 x 16 & ReLU \\
        Maxpool & 2 & - \\
        Linear & 16 x 6 & ReLU \\ 
        Linear & 120 x 84 & ReLU \\ 
        Linear & 84 x 10 & ReLU \\ 
        \hline
    \end{tabular}
\end{table}

\begin{figure}[hb]
    \centering
     \begin{subfigure}[b]{\textwidth}
    \centering
        \includegraphics[width=0.45\textwidth]{images/Fig4_legend.pdf}
    \end{subfigure}  
            \begin{subfigure}[b]{0.26\textwidth}
        \includegraphics[trim={0 0 0.4cm 0}, clip, width=\textwidth]{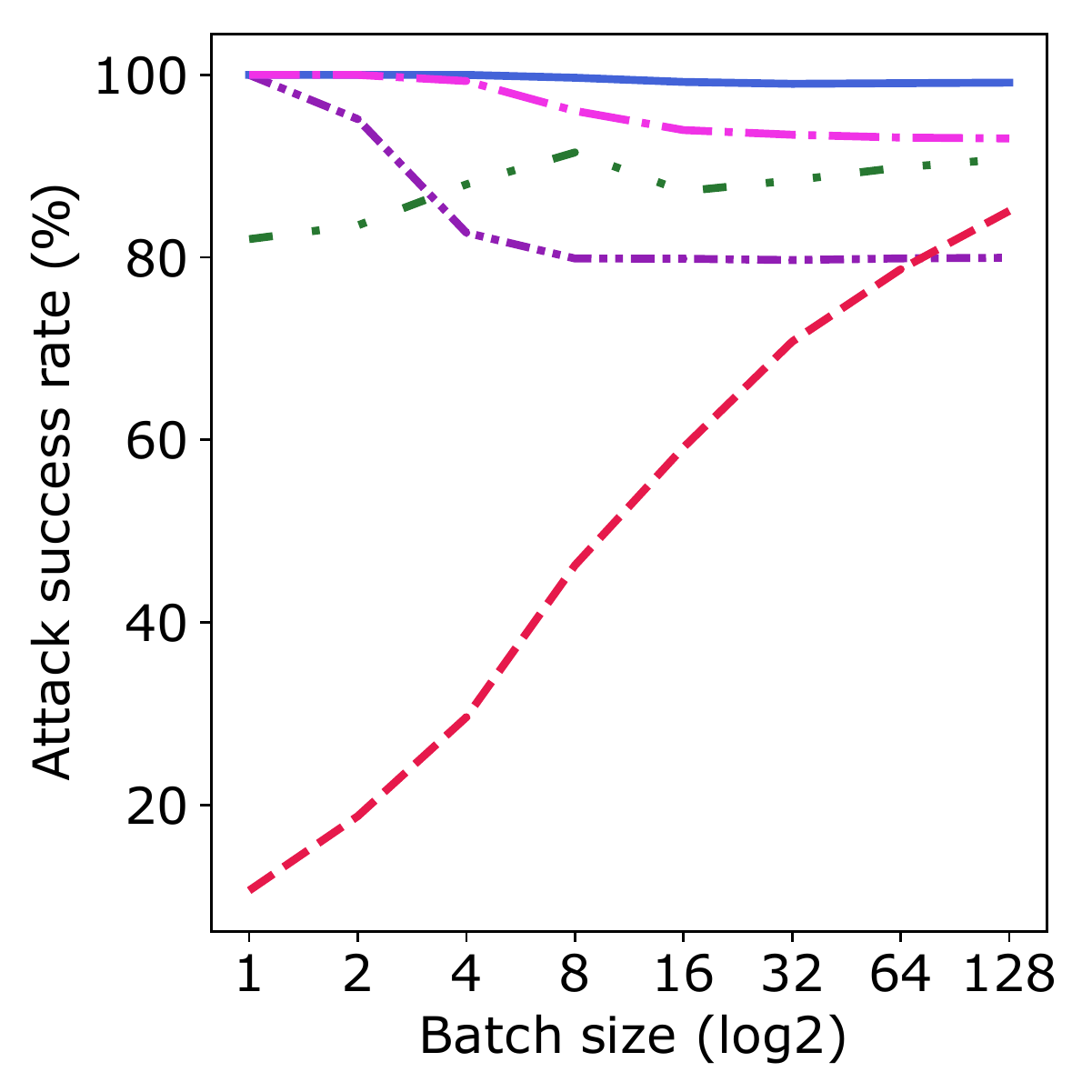} 
        \caption{MNIST - FS} 
        \label{fig:mnist_accuracy_iid}
    \end{subfigure}   
    \begin{subfigure}[b]{0.235\textwidth}
        \includegraphics[trim={1.1cm 0 0.4cm 0}, clip, width=\textwidth]{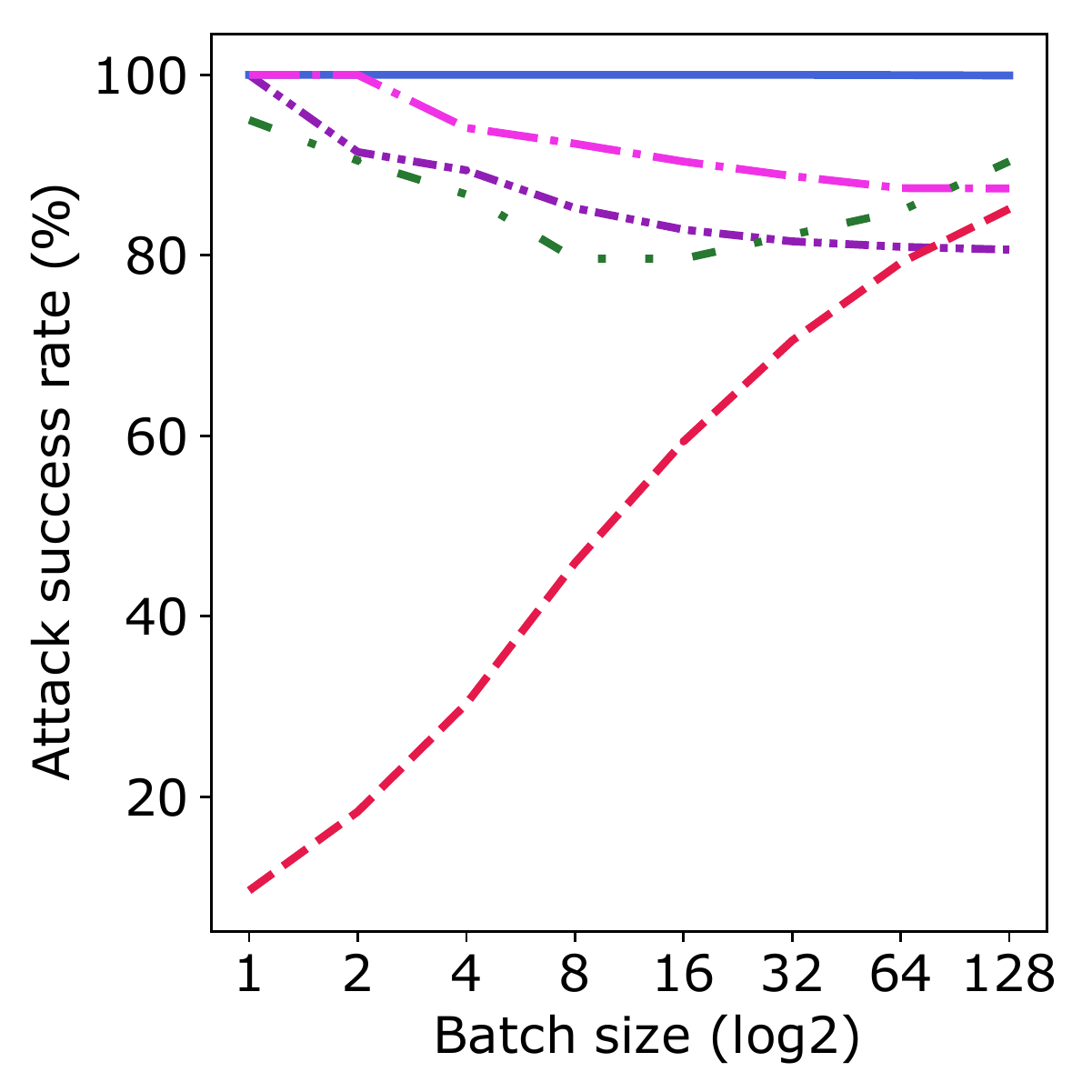} 
        \caption{SVHN - FS} 
        \label{fig:svhn_accuracy_iid}
    \end{subfigure} 
    \begin{subfigure}[b]{0.235\textwidth}
        \includegraphics[trim={1.1cm 0 0.4cm 0}, clip, width=\textwidth]{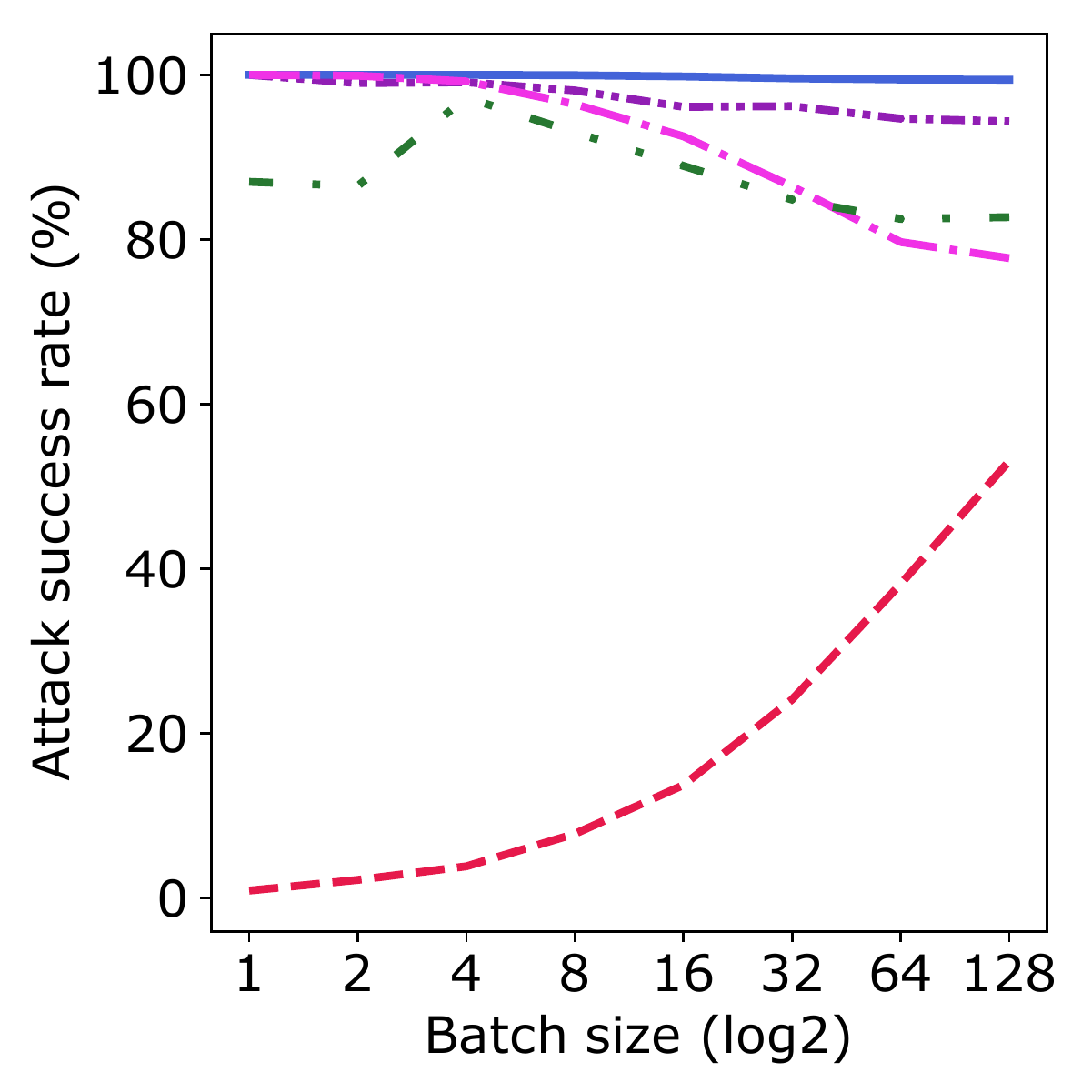} 
        \caption{CIFAR-100 - FS} 
        \label{fig:cifar_accuracy_iid}
    \end{subfigure}  
    \begin{subfigure}[b]{0.235\textwidth}
        \includegraphics[trim={1.1cm 0 0.4cm 0}, clip, width=\textwidth]{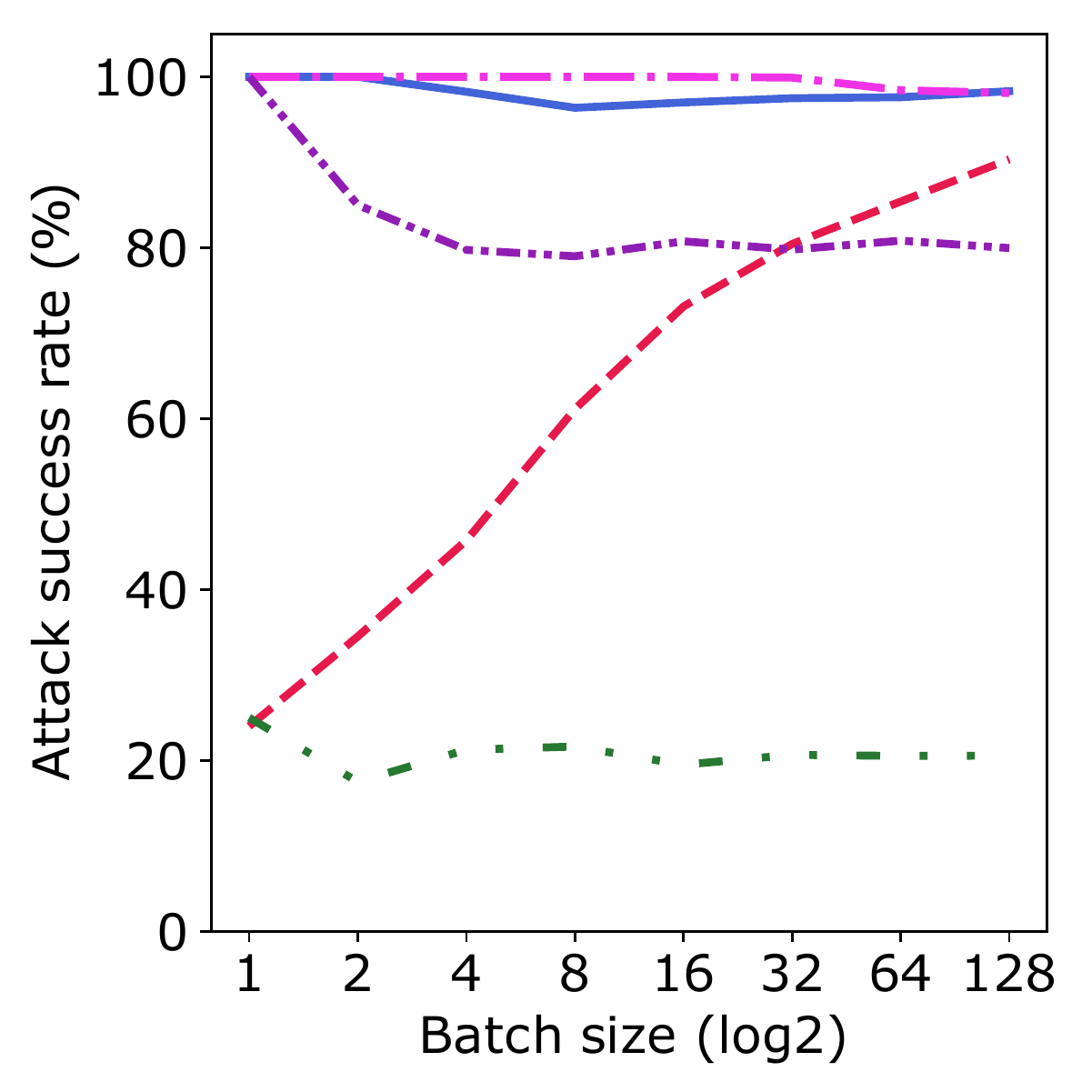} 
        \caption{CelebA - FS} 
        \label{fig:celebA_accuracy_iid}
    \end{subfigure}

    \caption{\injj{\footnotesize Attack success rate of (1) LLG with shared gradients, (2) LLG* with white-box model, (3) LLG+ with auxiliary knowledge, (4) DLG~\cite{zhu2019deep}, and (5) random guess on MNIST, SVHN, CIFAR-100, and CelebA. 
    Label extraction is based on gradients generated from passing one \emph{balanced} batch (FedSGD) to a randomly initialized CNN.
    }}
    \label{fig:grad_acc_balanced}
\end{figure}